\documentclass[twocolumn]{aastex631}

\newcommand{\elias}{K21}
\newcommand{\palio}{P20}
\newcommand{\uvot}{UV/optical}

\usepackage{amsmath}

\shorttitle{AGN \uvot\ power spectra}    
\shortauthors{Panagiotou et al.}

\graphicspath{{}{/home/pucax/mit/agn/snf_project/plots/}}

\begin{document}

\title{A physical model for the UV/optical power spectra of AGN}

\author{Christos Panagiotou}
\affiliation{MIT Kavli Institute for Astrophysics and Space Research, Massachusetts Institute of Technology, Cambridge, MA 02139, USA}

\author{Iossif Papadakis}
\affiliation{Department of Physics and Institute of Theoretical and Computational Physics, University of Crete, 71003 Heraklion, Greece}
\affiliation{Institute of Astrophysics, FORTH, Voutes, GR-7110 Heraklion, Greece}

\author{Erin Kara}
\affiliation{MIT Kavli Institute for Astrophysics and Space Research, Massachusetts Institute of Technology, Cambridge, MA 02139, USA}

\author{Elias Kammoun}
\affiliation{IRAP, Universit\'e de Toulouse, CNRS, UPS, CNES, 9, Avenue du Colonel Roche, BP 44346, F-31028, Toulouse Cedex 4, France}
\affiliation{INAF -- Osservatorio Astrofisico di Arcetri, Largo Enrico Fermi 5, I-50125 Firenze, Italy}

\author{Michal Dov\v ciak}
\affiliation{Astronomical Institute of the Academy of Sciences, Bo\v cn\'i II 1401, CZ-14100 Prague, Czech Republic}

\begin{abstract}


The \uvot\ variability of AGN has long been thought to be driven by the X-ray illumination of the accretion disk. However, recent multi-wavelength campaigns of nearby Seyfert galaxies seem to challenge this paradigm, with an apparent discrepancy between observations and the underlying theory. In order to further probe the connection between the \uvot\ and X-ray variability in AGN we developed a physical model to reproduce the \uvot\ power spectra (PSDs) of AGN assuming the thermal reprocessing of the X-rays in the disk. This model offers a novel way to probe the innermost regions of AGN. We use our model to study the variability of NGC 5548 and we infer that the X-ray and \uvot\ PSDs as well as the interband \uvot\ time lags are all well reproduced. We also derive constraints on the source physical parameters, such as the X-ray corona height and the accretion rate. Our results suggest that X-ray disk reprocessing accounts for the full variability properties of this AGN, within the considered time scales. Using earlier data of NGC 5548, we also show that our model can reproduce its PSD in different epochs, establishing the feasibility of using PSD modelling to investigate the time evolution of a source.

\end{abstract}

\section{Introduction} 
\label{sec:intro}

The early idea of matter accretion onto a massive compact object serving as the required energy reservoir for Active Galactic Nuclei \cite[AGN, e.g.][]{1964ApJ...140..796S} is nowadays widely accepted and has proven able to explain several observational characteristics of these objects \citep[e.g.][]{1999PASP..111..661S}. According to the contemporary view, surrounding matter accretes onto a supermassive black hole, which lies in the center of the nucleus, in the form of an optically thick and geometrically thin disk \citep{1973A&A....24..337S}. As the matter spirals inwards, part of the liberated gravitational energy heats the disk, which in turn emits quasi-thermal emission, typically referred to as disk black body radiation.

For typical AGN, the bulk of the disk radiation is preferentially emitted in the ultraviolet and optical part of the electromagnetic spectrum (\uvot, hereafter). AGN are known to be bright X-ray sources as well \citep{1978MNRAS.183..129E}. The X-ray emission is thought to be produced as disk photons are scattered to X-rays via inverse Compton scattering by high-energy electrons residing in the so-called X-ray corona \citep[e.g.][]{1993ApJ...413..507H}. While a few independent results suggest a small spatial extension for the corona, its exact geometry and physics remain rather unknown and the subject of active research \citep[e.g.][]{2013ApJ...769L...7R}. 

A long standing question in the study of AGN is how this X-ray source is connected to the accretion disk. Manifestations of the disk/corona interaction have, for example, been observed in the X-ray spectrum of AGN. Prominent spectral features, such as the Fe K$\alpha$ emission at 6.4 keV and the Compton hump at hard X-rays above 10 keV, are typically associated with X-ray reflection by the disk. The accretion disk has been unambiguously identified as the origin of these features in a few AGN, for which the variability of these emissions was found to lag shortly behind the variability of the X-ray continuum \citep[e.g.][]{2014ApJ...789...56Z, 2015MNRAS.446..737K}. More importantly, recent studies have modelled the observed lags in order to constain the physical properties of AGN \citep[e.g.][]{2018MNRAS.480.2650C, 2020MNRAS.498.4971M}.

Furthermore, if the X-rays illuminate the accretion disk, this should also be imprinted in the \uvot\ emission of AGN, where most of the disk radiation is observed. In fact, the absorption of X-rays will modify the effective temperature and the structure of the disk, and thus, the observed \uvot\ emission. As a result, we expect the X-ray and \uvot\ emission of AGN to be strongly connected, an expectation that was consistent with early studies \citep[e.g.][]{1992ApJ...393..113C}.

Nevertheless, it was not until recently that long and high-cadence multi-wavelength monitoring campaigns of nearby AGN made it possible to probe this connection in detail. For example, \cite{2015ApJ...806..129E} investigated the correlation between X-rays and \uvot\ in the case of NGC 5548, using \textit{Swift} and \textit{HST} observations. These authors found a rather low correlation between the two emissions and an interband time lag between the \uvot\ wavebands that was a factor of around 2-3 larger than expected in the case of a typical accretion disk, although the measured time lag was inferred to scale with the wavelength as expected, i.e. $\tau \propto \lambda ^{4/3}$. Similar results were then derived by later studies \citep[e.g.][]{2018MNRAS.480.2881M, 2019ApJ...870..123E}, which has led to the exploration of alternative mechanisms for the observed \uvot\ variability \citep[e.g.][]{2020ApJ...891..178S}.

On the other hand, \cite{2019ApJ...879L..24K} showed that the measured time lags for NGC 5548 are consistent with thermal reprocessing of X-rays by the disk, while \citet[][\palio ]{2020MNRAS.499.1998P} concluded that the observed \uvot\ variability amplitudes of this source are well explained as well, under the assumption of X-ray disk illumination. Finally, \cite{2021arXiv211001249D} showed that the disk reprocessing of X-rays can also reproduce the average spectral energy distribution of NGC 5548. 

Motivated by their previous results, \citet[][\elias]{2021ApJ...907...20K} performed an in-depth investigation of the disk emission when illuminated by X-rays and they derived an analytical model for the \uvot\ time lags as a function of wavelength, which was found to reproduce well the observed time lags in a number of sources \citep{2021MNRAS.503.4163K}. Using the model disk response of \elias, we study the expected \uvot\ variability of AGN in connection to the observed X-ray variability, under the assumption of X-ray illumination of the accretion disk. Understanding the relation between the \uvot\ and X-ray variability of AGN is of significant importance, as the characteristics of this relation would allow us to probe the nature and geometry of AGN. 

It is customary to investigate the connection between the \uvot\ and X-ray variability in the time domain, by studying their cross-correlation and the corresponding time lags \citep[e.g.][]{2019ApJ...870..123E}. Since the measured time lags depend predominantly on the temporal width of the disk response (\elias), their values encode only part of the information regarding the disk/corona interplay. Additional and independent constraints are provided by the variability amplitudes of the X-ray and \uvot\ emission, which depend on both the shape and the amplitude of the disk response. Further, and as shown in the following section, the study of the variability amplitudes is advantageously conducted in the Fourier space, which is the approach we employ in the present work. In particular, we investigate the expected shape of the \uvot\ power spectral distribution (PSD), which quantifies the variability of the \uvot\ light curve over different time scales, and we develop an analytical model to directly fit the PSD.

Several previous works have studied the \uvot\ PSDs of AGN \citep[e.g.][]{2011ApJ...743L..12M, 2014ApJ...788...33K}. However, these studies used a parametric and phenomenological model to reproduce the observed PSD, focusing mostly on its broad characteristics. Our aim is to use a physical model in order to directly test the assumption of AGN \uvot\ variability being driven by the X-ray illumination of the disk and to constrain the physical properties of AGN.

Moreover, we focus in probing the short term (of the order of days to months for a black hole mass of $10^7 $ M$_\odot$) variabilty of AGN. In longer time scales the X-ray illumination of the disk may not be the primary driver of the \uvot\ variability. For instance, \cite{2003ApJ...584L..53U} found that, while the X-ray light curve of NGC 5548 features a larger variability amplitude than its optical emission in short time scales, the opposite was true in longer time scales. This suggests that the reprocessing of the X-ray emission cannot be the sole explanation for the long term optical variability. In these longer time scales, other physical processes, such as the propagation of mass accretion fluctuations \citep[e.g.][]{2006MNRAS.367..801A}, may contribute to the observed \uvot\ variability. However, these processes are not expected to affect the short term disk variability due to their long characteristic time scales. In fact, in all the well studied AGN, the UV variations are larger and drive the optical variations \citep[see][for a recent review]{2021iSci...24j2557C}, i.e. the variability appears to propagate outwards in the disk, while the opposite trend is expected in the case of propagating accretion fluctuations. As a result, such processes seem to not be of relevance when the short term \uvot\ variability is studied.

This work is organized as follows. We present the computation of the expected PSD in Sect. \ref{sec:predict}, while we discuss its dependence on the various physical parameters in Sect. \ref{sec:phys_depend}. Section \ref{sec:5548} presents the application of the derived model to the observations of NGC 5548 and Sect. \ref{sec:discuss} discusses the findings of the present work. We summarize our main conclusions in Sect. \ref{sec:conclude}.

\section{Predicting the UV/optical PSD}
\label{sec:predict}

\subsection{The transfer function}
\label{sec:bkg}

Let us consider the case of the accretion disk being illuminated by the X-ray source. Then, part of the incident X-ray photons are absorbed by the disk, which increases the disk temperature and consequently, modifies the continuum disk emission. Since the photon absorption depends on, among others, the incident X-ray radiation, the observed disk emission is strongly connected to the X-ray emission.

In mathematical terms, this connection between the disk and X-ray flux can be expressed via the following equation:

\begin{equation}
\label{eq:convolution}
    F_\lambda (t) = F_\mathrm{NT}(\lambda) + \int_0^\infty \psi_\lambda(t') \cdot F_X (t-t') dt',
\end{equation}

\noindent where $F_\lambda$ denotes the total disk flux at a given wavelength $\lambda$, $F_\mathrm{NT}(\lambda)$ is the emission of a standard accretion disk in the case of no X-ray illumination \citep{1973blho.conf..343N}, $F_X$ denotes the X-ray flux, and $\psi$ is the so-called response function, which characterizes the disk emission (or response) due to illumination by an X-ray flash. Equation \ref{eq:convolution} describes the connection between the disk and the X-ray emission in the time domain. At any given time $t$, the disk flux depends on past values of the X-ray flux, with the specifics of this dependence being encoded in the shape of the response function. 

The connection between the disk and primary X-ray emission is significantly simplified, at least in mathematical terms, when considered in the Fourier space; and, more precisely, when one considers the corresponding variability power spectra. When eq. \ref{eq:convolution} holds, it is straightforward to show that \citep[e.g.][]{nla.cat-vn2888327,2016A&A...588A..13P}:

\begin{equation}
\label{eq:psd_conv}
    PSD_\lambda (\nu) = |\Gamma_\lambda(\nu)|^2 \cdot PSD_X (\nu),
\end{equation}

\noindent where $PSD_\lambda$ denotes the power spectrum of the disk light curve at wavelength $\lambda$, $PSD_X$ denotes the power spectrum of the X-ray light curve, and $\Gamma$ is the so-called transfer function, defined as the Fourier transform of the response function:

\begin{equation}
\label{eq:transfer}
    \Gamma_\lambda(\nu) = \int_{-\infty}^{\infty} \psi_\lambda(t) e^{-2\pi i\nu t} dt.
\end{equation}

It is worth highlighting that by studying the connection between the X-ray and disk emission in the Fourier space the problem is simplified from evaluating a convolution integral over time (eq. \ref{eq:convolution}) to computing a single multiplication (eq. \ref{eq:psd_conv}). 

Under the assumption of a steady system\footnote{As has been noted by a few recent studies, the assumption of a steady system is not necessarily valid in the case of AGN (e.g. \palio). We further discuss how this affects our analysis in Sect. \ref{sec:dynamic_evol}}, that is a system with a constant geometric and physical configuration over time, the transfer function, $\Gamma$, remains the same over time and depends solely on the physical parameters and geometry of the accretion disk/X-ray source system. In consequence, by estimating $\Gamma$ for different configurations of the system, one effectively produces a model to describe the connection between the UV/optical and X-ray light curves. When compared to observations (i.e. power spectra in our case), this model can be used to constrain the physical properties of AGN.

\subsection{Physical interpretation of $|\Gamma(\nu)|^2$}
\label{sec:phys_interp}

Ideally, one would use eq. \ref{eq:convolution} to study how the details of the X-ray variability are imprinted in the observed \uvot\ light curves. In practice, however, such an analysis is challenging due to the complex nature of the convolution integral and the rapid variability of the X-ray source. Instead, we aim to use eq. \ref{eq:psd_conv} in order to study the aforementioned relation.

Our objective is to estimate the transfer function for various model parameters, and use them to reproduce the observed \uvot\ PSDs, given the X-ray PSD. To that extent, we used the response functions derived by \elias\ (Sect. \ref{sec:transfer}). These authors defined the response function so that the observed disk flux at a given wavelength $\lambda$ is given by:

\begin{equation}
\label{eq:disk_elias}
    F_{\lambda,\text{obs}} (t) = F_\mathrm{NT}(\lambda) + \int_0^\infty L_{\text{Xobs,Edd}} (t-t') \cdot \psi_\lambda(t')  dt',
\end{equation}

\noindent where $L_{\text{Xobs,Edd}}$ is the observed X-ray (2-10 keV) luminosity measured in units of the Eddington luminosity, $L_{\text{Edd}}$. The response function was normalized over the X-ray luminosity so that the integral on the right hand side of the above equation gives the variable disk flux due to X-ray reprocessing in flux units (i.e., erg/s/cm$^2$). 

The power spectra of the observed X-ray and \uvot\ light curves are related as in eq. \ref{eq:psd_conv}. Since the power spectrum of a given light curve measures its variance density per unit frequency, then:

\begin{equation}
\label{eq:psd_variance}
\begin{split}
    \sigma^2_{\lambda \text{,obs}} & = \int_0^\infty  PSD_{\lambda \text{,obs}} (\nu) d\nu \\ 
                & = \int_0^\infty |\Gamma_\lambda(\nu)|^2 \cdot PSD_{\text{Xobs,Edd}} (\nu) d\nu,
\end{split}
\end{equation}

\noindent where $\sigma^2_{\lambda \text{,obs}}$ is the variance of the observed \uvot\ light curve. In the above equation, $PSD_{\text{Xobs,Edd}}$ is the power spectrum of the X-ray light curve normalized to its average flux (in units of $L_{\text{Edd}}$), and as a result, it is given in units of frequency$^{-1}$. On the contrary, being the power spectrum of the un-normalized \uvot\ light curve, $PSD_{\lambda \text{,obs}}$ quantifies the absolute disk variability amplitude and hence, it is given in physical units, which are set by $|\Gamma_\lambda(\nu)|^2$.

The physical interpretation and importance of $|\Gamma(\nu)|^2$ become evident from eq. \ref{eq:psd_variance}. For a given X-ray PSD, $|\Gamma(\nu)|^2$ determines the contribution to the total variance of the \uvot\ light curve by the Fourier components with frequency between $\nu$ and $\nu+d\nu$. In other words, the shape and amplitude of $|\Gamma(\nu)|^2$ determines the expected disk variability in different time scales.

In the following, we present how we estimated the transfer function for various system parameters and discuss its broad characteristics. We also derive an analytical approximation for the transfer function, which will then be applied to observational data.

\subsection{Computation of the transfer function}
\label{sec:transfer}

The reprocessing of X-rays by the disk depends, in a rather non-trivial way, on the specific characteristics of the system, such as the ionization profile of the disk and the incident X-ray radiation. Using the \texttt{KYNXILREV} model, \elias\ conducted an in-depth study of the disk emission due to X-ray illumination, taking into account all the relativistic effects. These authors assumed a lamp post geometry for the X-ray corona\footnote{Under the lamp post assumption, the X-ray source is considered to be a point source located along the spin axis of the black hole, which is also aligned with the disk's axis of rotation.} and a plane accretion disk reaching the innermost stable circular orbit. They further assumed a color correction factor of 2.4 \citep{1992MNRAS.258..189R} and they computed the response of the disk when illuminated by a short X-ray flash with a total temporal width of $T_\text{flash} = 10 ~t_\text{g}$, $t_\text{g}$ being the gravitational time\footnote{The gravitational time is defined as $t_\text{g} = \frac{R_\text{g}}{c} = \frac{GM_\text{BH}}{c^3}$, where $R_\text{g}$ denotes the gravitational radius, $G$ is the gravitational constant and $M_\text{BH}$ is the black hole mass. }. They calculated the disk response function for various physical parameters and several \uvot\ wavebands, and they studied how the disk response depends on them. 

Following eq. \ref{eq:transfer}, we used the response functions derived by \elias\ and we estimated the corresponding transfer functions as their discrete Fourier transform:

\begin{equation}
\label{eq:transfer_compute}
    \Gamma_\lambda[\nu_j] = \frac{\sum_{k=0}^{N-1} \psi[t_k] e^{-2\pi i \nu_j t_k} \Delta t}{\Delta \lambda \cdot e^{i\pi \nu_j T_\text{flash}} \text{sinc}(\pi \nu_j T_\text{flash})},
\end{equation}

\noindent where $N$ and $\Delta t$  are the total number of points and the bin size of the response function, respectively. The transfer function is calculated at the discrete frequencies $\nu_j = \frac{j}{N \cdot \Delta t}$, where $j=1, 2,...,N/2$, while the response function has been calculated in the discrete time points $t_k = k \cdot \Delta t$, where $k=0, 1,...,N-1$. In estimating the transfer function, the Fourier transform of the response function is divided by the width, $\Delta \lambda$, of the corresponding waveband in order to obtain $|\Gamma_{\lambda}[\nu_j]|$ in the necessary units of erg/s/cm$^2$/\AA. We further divided by the factor\footnote{Here, the sinc function is defined as $\text{sinc}x = \frac{\text{sin}x}{x}$.} $e^{i\pi \nu T_\text{flash}} \text{sinc}(\pi \nu T_\text{flash})$ in order to account for the finite width of the X-ray flash \citep[see discussion in Appendix C of][]{2016A&A...594A..71E}.

The transfer function was computed for the same values of the X-ray corona height, $h_\text{X}$, the black hole mass, $M_\text{8}$\footnote{$M_\text{8}$ denotes the black hole mass measured in units of $10^8 M_\odot$.}, the accretion rate, $\dot{m}$, the X-ray luminosity, $L_\text{X}$, the X-ray spectrum photon index, $\Gamma_\text{X}$, and the disk inclination, $\theta$\footnote{The inclination, measured in degrees, is defined as the angle between our line of sight and the disk's unit normal vector.} as those considered by \elias\ (Table 1 in that paper). Following \elias, we also estimated two sets of transfer functions; for a non-spinning ($\alpha=0$) and for a maximally spinning ($\alpha=1$) black hole, respectively. In the latter case, prograde accretion was assumed. In addition, for each set of physical parameters, the transfer function was computed for seven \uvot\ wavebands, which are representative of the wavebands considered in recent monitoring campaigns. Their width and central wavelength are listed in Table \ref{tab:wavebands_details}.

\begin{table}[]
    \centering
    \caption{Central wavelengths and widths of the wavebands for which we computed transfer function.}
    \begin{tabular}{lll}
    \hline
    Name & $\lambda_0 $(\AA)  & $\Delta \lambda $(\AA)   \\     
    \hline
    H1'      & 1158     & 5      \\
    H3'      & 1746     & 5      \\
    W2'      & 1950     & 600     \\
    W1'      & 2600     & 700     \\
    u'       & 3580     & 339     \\
    g'       & 4754     & 1387     \\
    r'       & 6204     & 1240     \\
    \tableline 
    \end{tabular}
    \label{tab:wavebands_details}
\end{table}

\begin{figure}[]
\includegraphics[width=\linewidth,height=1.1\linewidth, trim={0 0 0 0}, clip]{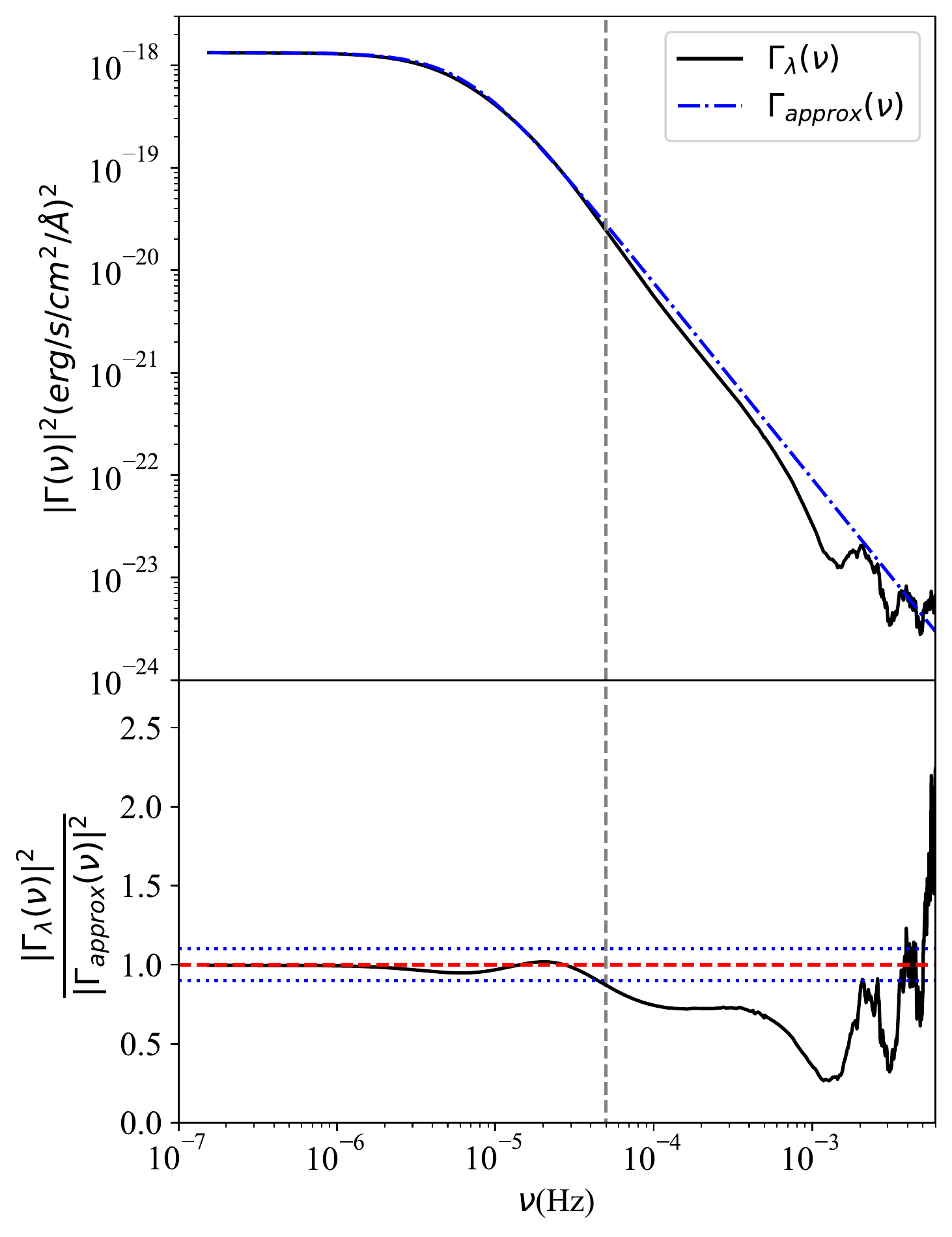}
\caption{\textit{Upper}: Plot of $|\Gamma_{mod}(\nu)|^2$ for the W2' band and for $h_\text{X} = 10~R_\text{g}$, $M_\text{8}=0.1$, $\dot{m} = 0.05~\dot{m}_\text{Edd}$, $L_\text{X} = 0.01 ~L_\text{Edd}$, $\Gamma_\text{X}=2.0$, and $\theta = 40 ^\circ$ (black solid line). The dot-dashed blue line denotes $|\Gamma_{approx}(\nu)|^2$ as defined by eq. \ref{eq:transfer_benpl_main}. The vertical grey dashed line denotes $\nu_\text{lim}$ (see Sect. \ref{sec:approximation}). The plotted functions were smoothed for clarity reasons. \textit{Lower}: The ratio between the model and the approximated $|\Gamma(\nu)|^2$. The red dashed line denotes the $y=1$ line, while the blue dotted lines denote the $\pm$10\% deviation from that line. } 
\label{fig:transfer_fidvals}
\end{figure}

As an example, Fig. \ref{fig:transfer_fidvals} shows the squared norm of the transfer function for the W2' waveband and the fiducial values of \elias\ ($h_\text{X} = 10 ~R_\text{g}$, $M_\text{8}=0.1$, $\dot{m} = 0.05~\dot{m}_\text{Edd}$, $L_\text{X} = 0.01 ~L_\text{Edd}$, $\Gamma_\text{X}=2.0$, and $\theta = 40 ^\circ$). As already discussed in Sect. \ref{sec:phys_interp}, for a given X-ray power spectrum, $|\Gamma_{\lambda}(\nu)|^2$ depicts the disk variability amplitude at each frequency. Consequently, a few interesting remarks about the expected disk variability over different time scales may already be made from Fig. \ref{fig:transfer_fidvals}.

The squared modulus of the transfer function forms a plateau at low temporal frequencies, remaining more or less constant. At this frequency range, or equivalently at long time scales, the whole accretion disk has already responded to the X-ray variations. As a result, the \uvot\ light curve will feature a long term variability similar to that of X-rays, with an amplitude set by $|\Gamma_{\lambda}(\nu)|^2$, which in turn depends on the amplitude of the response function and should be wavelength dependent.

The low-frequency plateau extends until a characteristic frequency, say $\nu_b$, which depends primarily on the time needed for the X-rays to illuminate the entire disk area responsible for the emission at a specific wavelength, as seen by the observer; that is to say, $\nu_b$ depends on the width of the response function. 

At higher frequencies, or equivalently shorter time scales, $|\Gamma_{\lambda}(\nu)|^2$ decreases quickly to lower values, following a kind of wavy pattern around a power-law decline, until it reaches significantly smaller values over which it oscillates. The time scales that correspond to $\nu>\nu_b$ are shorter than the time needed for X-rays to illuminate the respective disk area. Since a smaller disk area responds to the fast X-ray variations, we expect the disk variability amplitude, and so $|\Gamma_{\lambda}(\nu)|^2$, to be smaller than the amplitude of the longer time scales variability. Similarly, we expect $|\Gamma_{\lambda}(\nu)|^2$ to decrease with increasing frequency. In other words, the X-ray reprocessing by the disk smooths out the high-frequency X-ray fluctuations, making the \uvot\ light curve less variable than the X-rays at high frequencies. The high-frequency wiggles in Fig. \ref{fig:transfer_fidvals} are the result of the Fourier transform of a sharply increasing function of finite width.

\begin{figure}[]
\includegraphics[width=\linewidth,height=0.65\linewidth, trim={0 0 0 0}, clip]{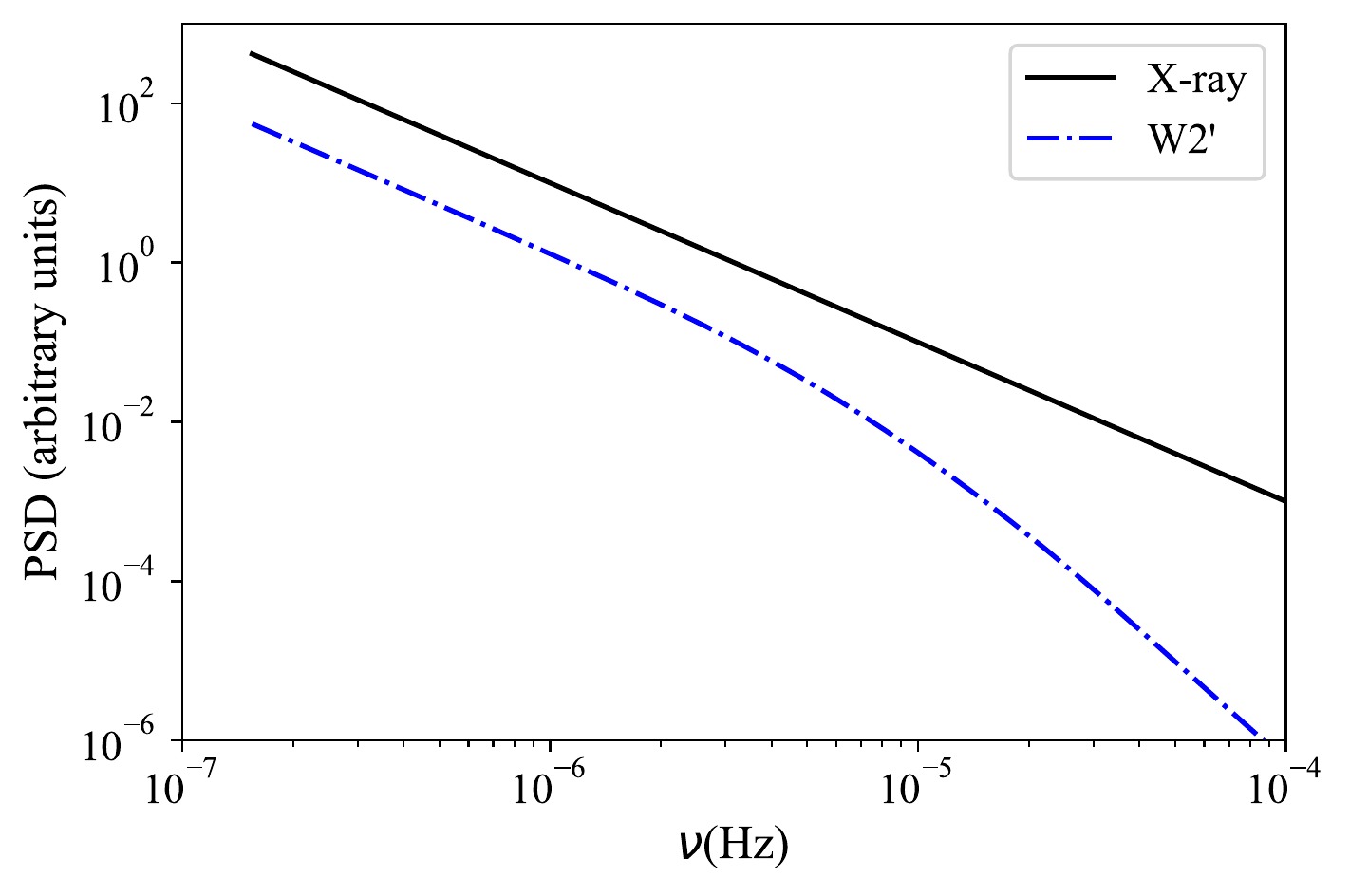}
\caption{Plot of an X-ray (black solid line) and W2' PSD (blue dot-dashed line). The X-ray PSD follows a power law with slope -2, while the W2' PSD is obtained using eq. \ref{eq:psd_conv} and $|\Gamma_{\lambda}(\nu)|^2$ of Fig. \ref{fig:transfer_fidvals}. The W2' PSD is shifted vertically to allow direct comparison to the X-ray PSD. } 
\label{fig:psd_fidvals}
\end{figure}


Following the above discussion, Fig. \ref{fig:psd_fidvals} plots the expected UV PSD using the transfer function of Fig. \ref{fig:transfer_fidvals}, and assuming an X-ray power spectrum following a power law with slope -2, which is also plotted in Fig. \ref{fig:psd_fidvals}. At low frequencies the UV PSD has a similar shape to that of the X-ray PSD, while at higher frequencies the UV PSD declines faster with frequency. A detailed discussion on the expected UV PSD of AGN given a generic X-ray PSD is given in Sect. \ref{sec:uvotpsd}.

\subsection{An analytical approximation for $|\Gamma(\nu)|^2$}
\label{sec:approximation}

The squared modulus of the transfer function as a function of frequency at all wavelengths was found to be well reproduced by a bending power law distribution with a zero slope at low frequencies, given by:

\begin{equation}
\label{eq:transfer_benpl_main}
    |\Gamma_{approx,\lambda}(\nu)|^2 = N  \cdot \frac{1}{1 +(\frac{\nu}{\nu_b})^{\large{s}}},
\end{equation}

\noindent where $s$ is the power-law slope at high frequencies, $\nu_b$ is the often called break or bending frequency, and $N$ denotes the normalization. The above function provides a good approximation for all the estimated transfer functions up to a certain frequency. The approximation was consistently poor at higher frequencies, where $|\Gamma_{\lambda}(\nu)|^2$ deviates from a power law (Fig. \ref{fig:transfer_fidvals}). Therefore, we used eq. \ref{eq:transfer_benpl_main} to model $|\Gamma_{\lambda}(\nu)|^2$ at frequencies $\nu < \nu_\text{lim}=5 \cdot 10^{-5}~ \text{Hz} = 4.32 ~\text{day}^{-1}$, at all wavelengths.

An example of the agreement between the model $|\Gamma_{\lambda}(\nu)|^2$ and $|\Gamma_{approx}(\nu)|^2$ is shown in the lower panel of Fig. \ref{fig:transfer_fidvals},  where we plot the ratio $|\Gamma_{\lambda}(\nu)|^2$ over $|\Gamma_{approx}(\nu)|^2$, as determined by fitting $|\Gamma_{\lambda}(\nu)|^2$ below $\nu_\text{lim}$.  Equation \ref{eq:transfer_benpl_main} describes sufficiently well the transfer function, with an accuracy of just a few percent, up to $\nu_\text{lim}$. At larger frequencies, $|\Gamma_{\lambda}(\nu)|^2$ cannot be well reproduced by our eq. \ref{eq:transfer_benpl_main}. It should be mentioned that from a practical perspective, the high-frequency shape of $|\Gamma(\nu)|^2$ is of less importance, as present and near future multiwavelength monitoring campaigns of AGN probe frequencies well below $\nu_\text{lim}$, which corresponds to a sampling rate of around 0.23 days. Therefore, our choice to constrain our analysis to $\nu \le \nu_\text{lim}$ does not affect the applicability of $|\Gamma_{approx}(\nu)|^2$ to current observational data.

The squared moduli of all the estimated transfer functions up to $\nu_\text{lim}$ for four wavebands and the corresponding ratios of $|\Gamma_{\lambda}(\nu)|^2$ over $|\Gamma_{approx}(\nu)|^2$ are plotted in Appendix \ref{sec:plots}. Evidently, the derived approximation yields an adequate description of $|\Gamma_{\lambda}(\nu)|^2$. The corresponding $s$, $\nu_b$, and $N$ change for different system configurations and wavebands. The dependence of $s$, $\nu_b$, and $N$ on the various physical parameters was found to be well described by the following equations:

\vspace{11pt}
$\begin{array}{clr}
    \nu_b(\lambda, \vec{\mu}) &= a_\nu (\vec{\mu}) \cdot \lambda_{1950}^{-b_\nu(\vec{\mu})} & \\
    s(\lambda, \vec{\mu})     &= a_s(\vec{\mu}) \cdot \lambda_{1950}^{-b_s(\vec{\mu})}    & \\
    N(\lambda, \vec{\mu})   &=\begin{cases}  a_{N1}(\vec{\mu}) \cdot \lambda_{1950}^{-b_{N1}(\vec{\mu})} + c_{N1}(\vec{\mu}) , & \mbox{if }  \lambda_{1950}<1 \\ a_{N2}(\vec{\mu}) \cdot \lambda_{1950}^{-b_{N2}(\vec{\mu})}, & \mbox{if }\lambda_{1950} \ge 1, \end{cases}
\end{array}
$
\vspace{11pt}

\noindent where $\lambda_{1950} = \frac{\lambda}{1950 \AA}$ and $\vec{\mu}$ denotes the state of the physical system, namely \mbox{$\vec{\mu}=(M_8, \dot{m}, h, L_X, \Gamma_X, \theta)$}. For brevity, the exact form of the above equations is presented in detail in Appendix \ref{sec:a1}.

An intuitive understanding of how $s$, $\nu_b$, and $N$, and hence the expected disk variability, depend on the various physical parameters may be obtained from the plots in Appendix \ref{sec:plots}. This dependence is qualitatively discussed in the following section.


\section{Dependence of the transfer function on physical parameters}
\label{sec:phys_depend}

\subsection{Dependence of $|\Gamma(\nu)|^2$ on $M_8$ and $\alpha$}
\label{sec:depend_mass_spin}

Figure \ref{fig:transfer_ratio_mass} plots $|\Gamma_{\lambda}(\nu)|^2$ as a function of frequency for different values of the black hole mass. For a specific waveband, the normalization of $|\Gamma_{\lambda}(\nu)|^2$ increases with the mass, while the bending frequency drops to lower values. These trends become clear if one considers how the accretion disk is modified for bigger values of the black hole mass. For the same accretion rate, the disk is larger (in physical size) and its temperature is lower for a larger black hole. As a result, the X-ray variations require more time to illuminate the whole disk, explaining in this way the decrease of $\nu_b$. On the other hand, the lower disk temperature suggests that a larger fraction of the observed disk emission would be the result of X-ray reprocessing. This implies a larger amplitude for the transfer function and, hence, a larger variability amplitude of the total emission (see also \elias\ for the dependence of the response amplitude on the black hole mass).

The calculated transfer functions depend on the spin of the central black hole as well. Since the transfer functions were estimated only for two values of the spin, this dependence cannot be investigated in detail. Nevertheless, it is worth noting here the main effects of the spin. On one hand, a higher spin results in the disk extending to a smaller inner radius, down to 1 $R_\text{g}$. This would increase the amount of X-ray emission being reprocessed by the disk, and as a result, the normalization of $|\Gamma_{\lambda}(\nu)|^2$ increases for larger spins. On the other hand, the spin also determines the value of the assumed accretion rate in physical units, as the radiative efficiency depends on the black hole spin. For the same accretion rate (measured as a fraction of $\dot{m}_\text{Edd}$), a lower spin corresponds to a physically larger amount of mass being accreted and subsequently, to a hotter disk. The bending frequency of $|\Gamma_{\lambda}(\nu)|^2$ is, therefore, expected to be slightly smaller, on average by a factor of 2 or even less, for slower rotating black holes.

\subsection{Dependence of $|\Gamma(\nu)|^2$ on $h_\text{X}$ and $L_\text{X}$}
\label{sec:depend_h_lx}

Figure \ref{fig:transfer_ratio_height} plots $|\Gamma_{\lambda}(\nu)|^2$ as a function of frequency for different values of the corona height,  $h_\text{X}$. This figure shows that for a specific waveband the normalization of $|\Gamma_{\lambda}(\nu)|^2$ increases with $h_\text{X}$. As  the corona height increases, the disk's solid angle, as seen by the X-ray source, increases as well and hence, the amount of X-ray emission being reprocessed in the disk is larger. As a result, the disk emission is expected to feature a larger variability amplitude when $h_\text{X}$ assumes higher values.

Moreover, Fig. \ref{fig:transfer_ratio_height} illustrates how the bending frequency, $\nu_b$, varies with $h_\text{X}$. For a given waveband, the bending frequency decreases when the corona height increases. As already discussed, the value of $\nu_b$ depends on the time required for the imposed X-ray variations to illuminate the entire disk area emitting at the corresponding waveband. As the distance of the X-ray source from the disk rises, this time increases accordingly, resulting in lower values for the bending frequency. It is worth noticing that for a given waveband the bending frequency depends only slightly on the height and the other physical parameters discussed below as the aforementioned time does not vary excessively with these parameters. For a specific physical parameter, $\nu_b$ varies at most by a factor of a few within the considered parameter space. 

The situation is a bit more puzzling when the dependence of $|\Gamma_{\lambda}(\nu)|^2$ on the X-ray luminosity is examined. The corresponding normalization and bending frequency decline as $L_\text{X}$ assumes larger values, which is depicted in Fig. \ref{fig:transfer_ratio_lumin}. As discussed in detail by \elias, the response function becomes broader for a larger X-ray luminosity and since $\nu_b$ depends on the response width, we expect $\nu_b$ to decrease with increasing $L_\text{X}$. The main reason for this decrease is that a larger X-ray luminosity increases the disk temperature, which results in the observed emission in a fixed bandpass to originate from a more extended region of the disk.

To explain the normalization trend, we should first point out that the employed response functions were normalized with respect to the assumed X-ray luminosity (see \elias, for details). Thus, the decrease of the transfer function's amplitude for larger X-ray luminosities merely implies that the variability of the reprocessed emission in absolute terms rises at a slower pace than $L_\text{X}$. In other words, $|\Gamma_{\lambda}(\nu)|$ does not scale linearly with $L_\text{X}$. This effect becomes evident for larger X-ray luminosities, while for $L_\text{X} < 0.01 \cdot L_\text{Edd}$ there are only minor differences in the amplitude of $|\Gamma_{\lambda}(\nu)|^2$ (Fig. \ref{fig:transfer_ratio_lumin}). As explained by \elias, the decrease of the amplitude of the response function, and equivalently of $|\Gamma_{\lambda}(\nu)|^2$, with increasing X-ray luminosity is due to the complex contribution of different disk segments to the observed emission and due to an increase in the disk's ionization degree for larger values of $L_\text{X}$, especially in its inner region.

\subsection{Dependence of $|\Gamma(\nu)|^2$ on $\dot{m}$ and $\Gamma_\text{X}$}
\label{sec:depend_mdot_gamma}

For a particular black hole mass, an increased accretion rate results in larger disk temperatures. Consequently, a smaller fraction of the observed emission is due to X-ray reprocessing, whereas the emission of a specific waveband originates from outer regions of the disk. These considerations explain why both the normalization and bending frequency of $|\Gamma_{\lambda}(\nu)|^2$ decrease for larger values of $\dot{m}$, as is illustrated in Fig. \ref{fig:transfer_ratio_mdot}.

Additionally, Fig. \ref{fig:transfer_ratio_gamma} plots $|\Gamma_{\lambda}(\nu)|^2$ for different values of the photon index, $\Gamma_\text{X}$. The photon index affects mainly the fraction of the X-ray emission being absorbed by the disk. In particular, for a given X-ray flux, $\Gamma_\text{X}$ determines the energy distribution of the emitted X-ray photons. A low $\Gamma_\text{X}$ (i.e. flat spectrum) corresponds to larger amount of high-energy photons, which are more likely to be reprocessed by the disk in the X-rays than in the \uvot, in comparison to low-energy X-rays. Accordingly, a low $\Gamma_\text{X}$ results in a smaller fraction of the X-ray emission being reprocessed and hence, in a cooler disk. This explains the increase of the transfer function's amplitude and the slight decline of its bending frequency in the case of higher $\Gamma_\text{X}$.

\subsection{Inclination dependence and the $|\Gamma(\nu)|^2$ slope}
\label{sec:depend_incl_slope}

The system's inclination angle influences $|\Gamma_{\lambda}(\nu)|^2$ in two ways. First, a higher inclination reduces the apparent area of the disk, lowering, thus, the observed disk emission. Second, the reprocessed emission on the near side of the accretion disk reaches the observer faster for higher inclination angles, while the reprocessed emission on the far side requires more time to reach the observer in this case. Thus, the disk emitting region appears to be more extended for higher inclination. The above considerations reveal that the normalization and the bending frequency of $|\Gamma_{\lambda}(\nu)|^2$ are expected to decrease as the source gets more inclined, exactly as demonstrated in Fig. \ref{fig:transfer_ratio_incl}. 

Moreover, Fig. \ref{fig:transfer_ratio_incl} shows that different values of the inclination angle predict distinctively different \uvot\ power spectra. As a result, the analysis of observed \uvot\ PSDs can, in principle, constrain the inclination of the source. This is in contrast to the commonly employed time lag spectra, which \elias\ have shown to be insensitive to variations of the inclination angle.

The above discussion has deliberately excluded the dependence of the high-frequency slope, $s$, on the various physical properties, which is the least intuitive. The slope quantifies the rate at which the variability amplitude increases from higher to lower frequencies, or equivalently from shorter to longer time scales. This rate of increase depends on the pace at which we observe the X-ray imposed variations to illuminate the different segments of the accretion disk. Therefore, and to a first approximation, the slope is expected to depend only marginally on the physical parameters, because this pace is rather insensitive to changes of the system configuration. Indeed, eq. \ref{eq:transfer_slope} and Figs. \ref{fig:transfer_ratio_mass}-\ref{fig:transfer_ratio_incl} confirm the mild variations of the slope as the different physical parameters vary. The only obvious exception is the inclination angle, which seems to modify significantly the high-frequency slope of $|\Gamma_{\lambda}(\nu)|^2$. This is, however, expected, since, as already explained, the inclination angle affects the apparent size and spatial extent of the disk, which alters profoundly the aforementioned pace.

\subsection{The wavelength dependence of $|\Gamma(\nu)|^2$}
\label{sec:depend_lamda}

Finally, for each system configuration the squared modulus of the transfer function depends on the considered waveband. In general, emission in shorter wavelengths originate primarily from inner regions of the accretion disk, which absorb larger fraction of the X-ray emission per unit area due to their increased solid angles as seen by the X-ray source. Therefore, the observed disk emission will be more variable for bluer wavebands. At the same time, since the emission in longer wavelengths originate from a more extended region of the disk, the bending frequency of $|\Gamma_{\lambda}(\nu)|^2$ will be lower in this case. Both of these results are illustrated in Figs. \ref{fig:transfer_ratio_mass}-\ref{fig:transfer_ratio_incl}.

\section{Application to observations: The case of NGC 5548}
\label{sec:5548}

\subsection{The use of $|\Gamma_{approx}(\nu)|^2$}
\label{sec:using_gamma_approx}

Equation \ref{eq:transfer_benpl_main} allows us to compute $|\Gamma(\nu)|^2$ for any system configuration and waveband, by interpolating within the used grid points. Consequently, we can apply eq. \ref{eq:transfer_benpl_main} to observational data to test if the observed \uvot\ variability is consistent with the assumption of X-ray thermal reprocessing. Moreover, using $|\Gamma_{approx}(\nu)|^2$ to fit the connection between \uvot\ and X-ray PSD, we can constrain the physical properties of the source.

In practice, both X-ray and \uvot\ PSDs should be computed by simultaneous observed light curves. In this case, the PSDs can be fitted jointly, using for example a phenomenological model for the X-ray PSD and eqs. \ref{eq:psd_conv}, \ref{eq:transfer_benpl_main} to model the \uvot\ PSDs. We performed such an analysis for the case of NGC 5548, which is presented in the following section.

Before discussing our results, it is worth mentioning that we examined whether the use of $|\Gamma_{approx}(\nu)|^2$, instead of the exact model estimated transfer function, introduces any biases to the results of the intended analysis. To that scope, we used $|\Gamma_{approx}(\nu)|^2$ to fit the model estimated $|\Gamma_{\lambda}(\nu)|^2$ of known physical parameters. As is shown in more detail in Appendix \ref{sec:a1}, this fit retrieves accurately the input physical parameters, which proves the validity of using eq. \ref{eq:transfer_benpl_main} to reproduce observed power spectra of AGN.

It should, further, be noted that the rather simple shape of $|\Gamma_{approx}(\nu)|^2$ does not allow the simultaneous estimation of all the model parameters by fitting a limited sample of PSDs due to the model degeneracy. For example, the X-ray source height is expected to be degenerate with the black hole mass, as they both affect $\nu_b$, $s$, and $N$ in a similar way.  As shown in Appendix \ref{sec:a1} though, the fit of five PSDs allows for the unbiased estimation of up to three model parameters. In this case, the other parameters may be constrained from independent information. For example, the X-ray luminosity and photon index may be fixed to the values obtained from the spectral analysis of the considered observations. Moreover, below we perform a joint fit of both the PSDs and the time lags, which should be preferred when possible, as it aids in breaking part of the degeneracies. 

\subsection{AGN STORM campaign}
\label{sec:storm_campaign}

NGC 5548 is a well-studied nearby AGN, located at redshift z=0.01717. This source was the target of a long multi-wavelength monitoring campaign in 2014 \citep[namely, the ``STORM" campaign,][]{2015ApJ...806..128D}. In \palio, we used the X-ray and \uvot\ light curves from that campaign in order to study the corresponding power spectra of this object. We found that all the considered PSDs can be reproduced well by a simple power-law model. Here, we examine whether the observed PSDs can be well reproduced by the more physical model presented in the previous sections.

The estimation of the power spectra was presented in detail in \palio. The X-ray and the \uvot\ PSDs were estimated using contemporaneous light curves. Therefore, we decided to fit both of them simultaneously, so that the parameters of the X-ray and \uvot\ PSD models were minimized in a combined fit.

We used the normalized (over the mean flux) X-ray PSD and the power spectra of the wavebands H1, H3, W2, W1, B, and V, as these are given in \palio. We converted the \uvot\ PSDs to physical units\footnote{As is customary, the periodograms in \palio\ were normalized to the mean flux of their light curves.} by multiplying them with the factor $(\bar{f}_\lambda - f_{\lambda,host})^2$, where $\bar{f}_\lambda$ is the average value of the corresponding light curve and $f_{\lambda,host}$ denotes the host galaxy contribution in each waveband, given in \cite{2016ApJ...821...56F}.

The hard X-ray\footnote{Defined in this energy range, the HX PSD is expected to not be significantly affected by any potential variability of the absorption in our line of sight. In fact and as shown below, all the \uvot\ PSDs are well reproduced given the observed HX PSD, which strongly suggests that the observed X-ray variability, although of unkown origin, is the same as the one reaching the accretion disk.} (2-7 keV, HX hereafter) power spectrum was modelled by a bending power law with a slope of -1 at low frequencies:

\begin{equation}
\label{eq:psd_xraymodel}
    P_{m,X} (\nu) = A_X \cdot \nu^{-1} \frac{1}{1 + (\frac{\nu}{\nu_{b,X}})^{s_x}} + C_X,
\end{equation}

\noindent which has been found to reproduce well the X-ray PSD of NGC 5548 over a large range of frequencies \citep{2003ApJ...593...96M}. The constant $C_X$ was added to account for the variability contribution due the experimental noise. 

Using eq. \ref{eq:psd_conv}, the \uvot\ PSDs were modelled as:

\begin{equation}
\label{eq:psd_uvmodel}
    P_{m,\lambda} (\nu) = f^2_{abs} \cdot f^2_{Gal} \cdot |\Gamma^\prime_{approx,\lambda}(\nu)|^2 P_{m,X} (\nu) \cdot L^2_\text{X} + C_\lambda,
\end{equation}

\noindent where $f_{abs}$ and $f_{Gal}$ are absorption factors used to model the effects of interstellar reddening in the host galaxy and our Galaxy, respectively\footnote{Assuming the absorption of the intrinsic emission, so that the observed flux is given by $F_\text{obs} = A \cdot F_\text{intr}$, where A is an absorption factor and $F_\text{intr}$ the intrinsic flux, it may be shown that the power spectrum of the observed emission is then equal to \mbox{$PSD_\text{obs} = A^2 \cdot PSD_\text{intr}$} \citep{2007ApJ...661...38P}.}. The former is given by \mbox{$f_{abs} = 10^{-A_\lambda/2.5}$}, $A_\lambda$ being the total extinction at wavelength $\lambda$, which was calculated using the reddening curves of \cite{1989ApJ...345..245C} and assuming $R_V=3.1$, while the corresponding $E(B-V)$ was left free to be determined during the fit. The Galactic extinction is calculated in the same way using further the known value $E(B-V) = 0.017$ \citep{2011ApJ...737..103S}.

$|\Gamma^\prime_{approx,\lambda}(\nu)|^2$ is equal to $|\Gamma_{approx,\lambda}(\nu)|^2$ of eq. \ref{eq:transfer_benpl_main} divided by $(\frac{D_\text{L}}{\text{Mpc}})^4$, where $D_\text{L} = 80.1 ~\text{Mpc}$ is the luminosity distance of NGC 5548, retrieved from \textit{NED}\footnote{https://ned.ipac.caltech.edu/}. This was necessary as the response functions were derived by \elias\ assuming a source distance of 1 Mpc. Finally, the X-ray PSD model is multiplied by $L^2_\text{X}$ so that it is given in normalized units of $L_\text{Edd}$ (see Sect. \ref{sec:phys_interp}). Similarly to the X-ray model, a constant, $C_\lambda$, was added to the model of \uvot\ power spectra to account for the experimental noise in the corresponding light curves and was left free during the fit.

The fit was performed in \textit{XSPEC} software version 12.12.0 \citep{1996ASPC..101...17A}, assuming Gaussian statistics. All the errors reported in this work correspond to 1-$\sigma$ confidence interval, unless otherwise noted.

\subsection{A joint fit of the PSDs and lag spectrum}
\label{sec:fit_psd_lags}

If the observed \uvot\ variability is mainly due to X-ray irradiation of the disk, then we should be able to fit, simultaneously, both the \uvot\ PSDs and the corresponding interband time lags observed in correlation studies. Such a combined fit will provide a much stronger constraint to the model, and in doing so, we will be able to estimate the model parameters more accurately. We therefore decided to perform a combined fit of both the X-ray, \uvot\ PSDs and the observed \uvot\ time lags estimated by \cite{2016ApJ...821...56F}. The time lag spectrum was fitted using the analytical model derived by \elias\ (eq. 8 in that paper). We considered the detected lags for the wavebands H1, H2, H3, W2, M2, W1, B, and V so that the time lag spectrum spans the same wavelength range as the considered PSDs.

During the fit, we kept the black hole mass fixed to $M_8=0.52$\footnote{Value retrieved from the AGN black hole mass catalog www.astro.gsu.edu/AGNmass \citep{2015PASP..127...67B}. The mass was estimated using the technique of reverberation mapping of broad optical lines \citep[e.g.][]{2004ApJ...613..682P} and assuming a scale factor $<f>=4.3$ \citep{2013ApJ...773...90G}. }. Following \cite{2021MNRAS.503.4163K}, the X-ray photon index and luminosity were also kept fixed to the values $\Gamma_\text{X}=1.5$ and $L_\text{X}=0.0034~L_\text{Edd}$, which were obtained by \cite{2017ApJ...846...55M} when considering the average X-ray spectrum of the source. Finally, we assumed $\theta=40^\circ$. The free parameters during the fit were the X-ray corona height, $h_\text{X}$, the accretion rate, $\dot{m}$, the host galaxy extinction, $E(B-V)$, and the model parameters of the X-ray PSD (eq. \ref{eq:psd_xraymodel}). We performed the fit twice; assuming a non-spinning and a maximally spinning black hole.

\begin{figure*}[]
\includegraphics[width=0.245\textwidth,height=0.45\textwidth, trim={0 0 0 0}, clip]{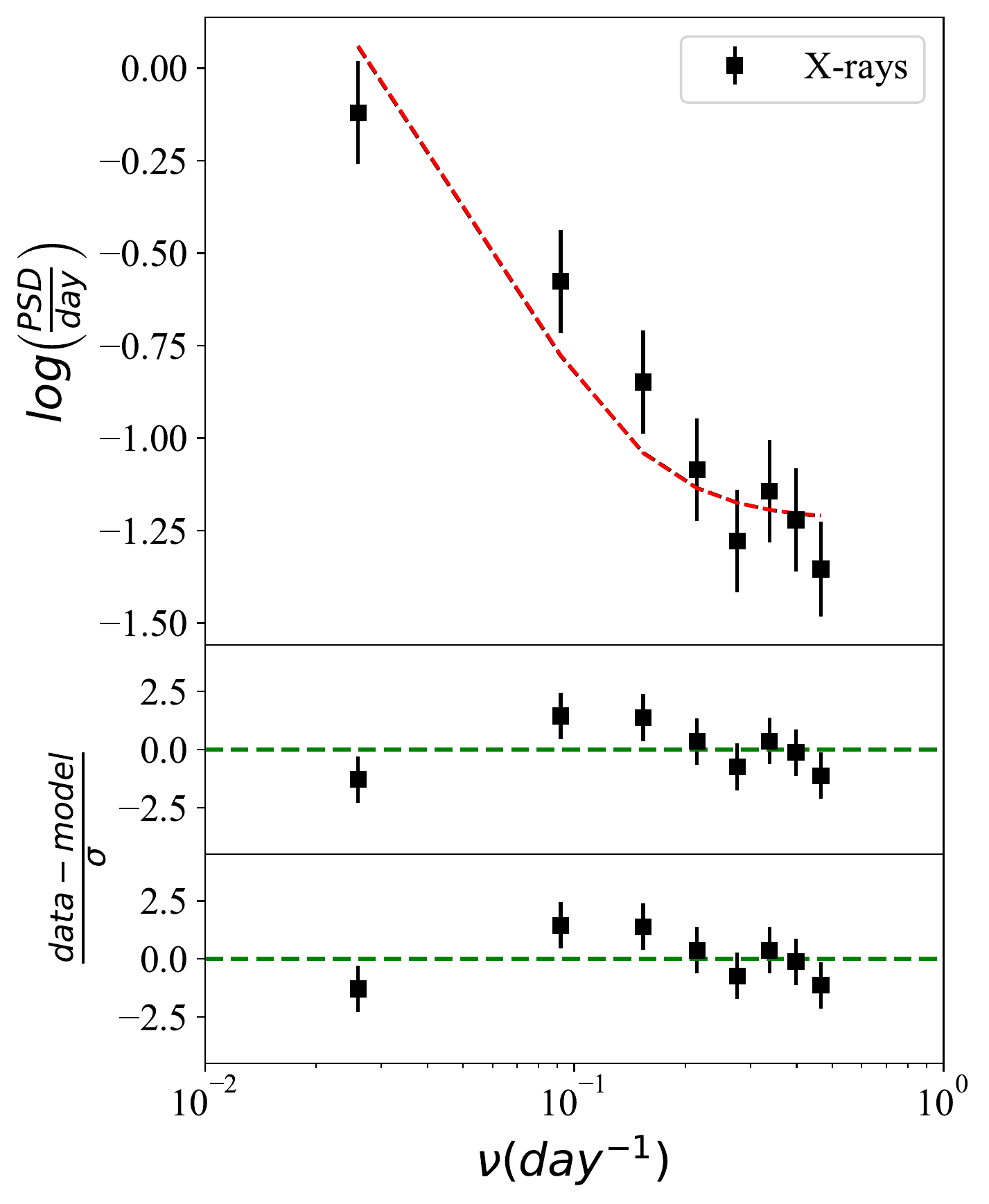}
\includegraphics[width=0.49\textwidth,height=0.45\textwidth, trim={0 0 0 0}, clip]{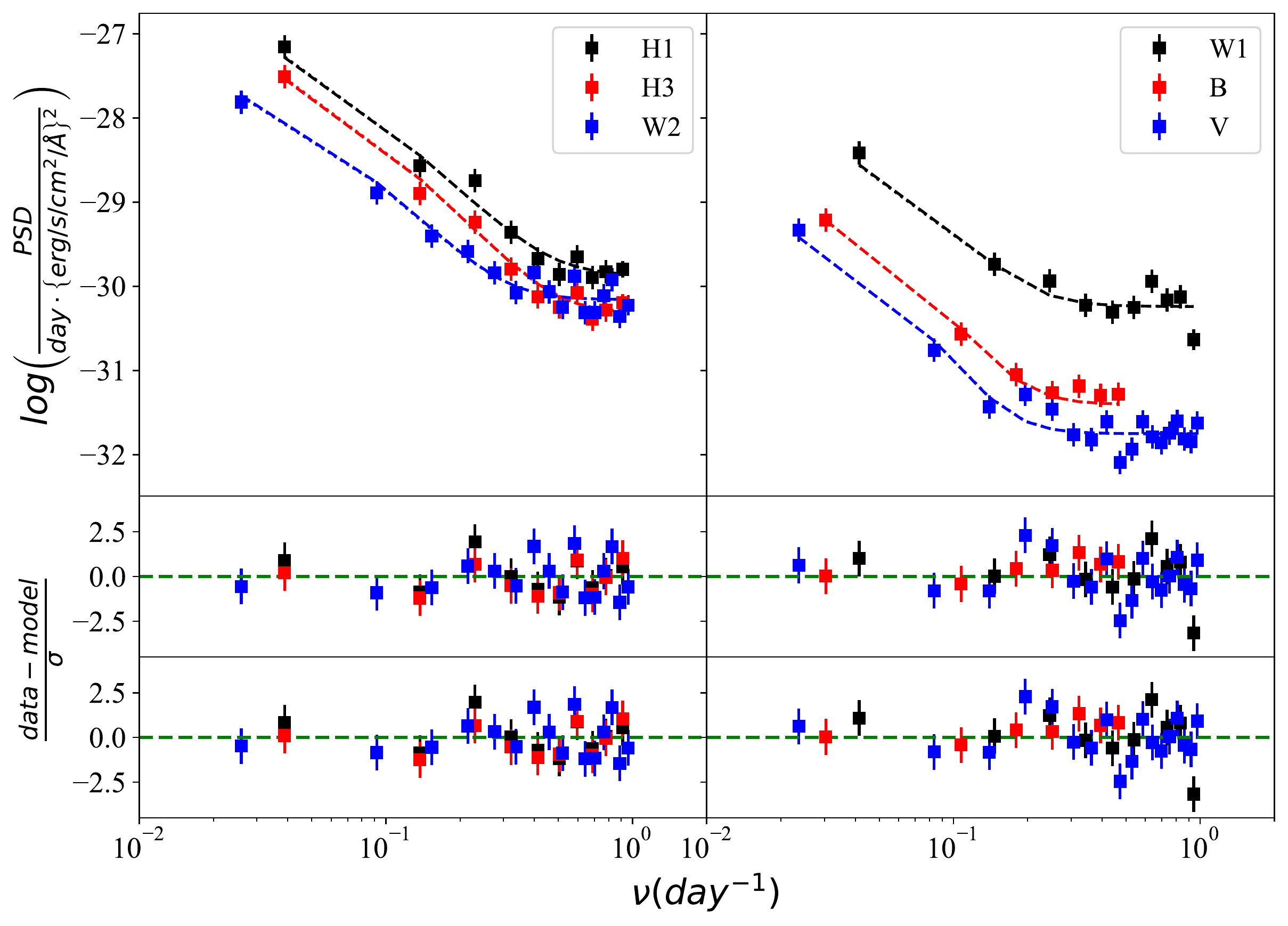}
\includegraphics[width=0.245\textwidth,height=0.45\textwidth, trim={0 0 0 0}, clip]{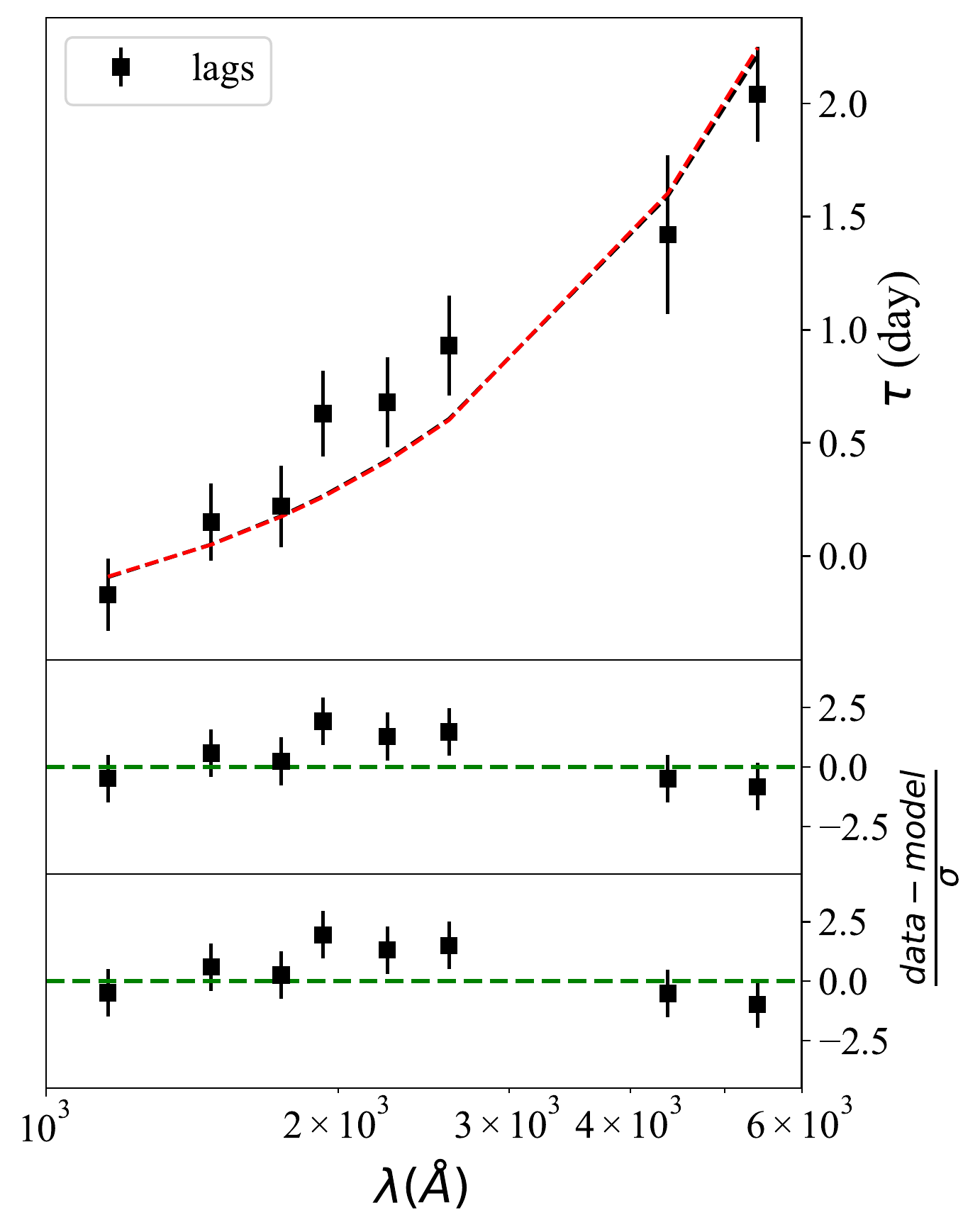}

\caption{\textit{Left}: X-ray PSD and the best-fit models for $\alpha=1$ (black) and $\alpha=0$ (red dashed line). \textit{Middle}: \uvot\ PSDs (squares) and the best-fit model (dashed line). For simplicity, only the model for maximum black hole spin is plotted.  \textit{Right}: \uvot\ time lags and the best-fit models for $\alpha=1$ (black) and $\alpha=0$ (red dashed line). The lower panels plot the fit residuals for $\alpha=1$ (middle) and $\alpha=0$ (bottom). } 
\label{fig:psdlags_bestfit}
\end{figure*}


The model fits well all the power spectra in the different wavebands as well as the measured time lags. The PSDs and the lag spectrum as well as the best-fit models are plotted in Fig. \ref{fig:psdlags_bestfit}. The lower panels of this figure show the fit residuals for both fits (i.e. $\alpha=0$ and $1$). The model agrees well with the data and no obvious trend is apparent in the residual plots. The quality of the fit was further confirmed by the deduced $\chi^2$ statistic (around 95 for 75 degrees of freedom or df), which corresponds to a chance probability of $P_\text{null} \simeq 5-6\%$ for both spins.

Table \ref{tab:bestfit_storm} lists the best-fit values for the model parameters. They are consistent, within the errors, between the spin 0 and 1 case, except for the best-fit accretion rate, which is significantly smaller in the $\alpha=0$ case. In fact, the best-fit value in this case is just an upper limit, as the lower limit of the confidence interval for $\dot{m}$ is equal to the lower value considered for this parameter by our model, that is $0.005~\dot{m}_\text{Edd}$.

The best-fit values for the X-ray PSD (i.e. the high-frequency slope, the break frequency, and the normalization) are consistent within the errors with the values reported by \citep{2003ApJ...593...96M}, suggesting a common X-ray variability behavior over many years.

\begin{table}[]
    \centering
    \caption{Best fit results when the \uvot\ and X-ray PSDs of the STORM campaign as well as the time lags are fit simultaneously.}
    \begin{tabular}{llll}
    \hline
    Parameter        &   Units     &  $\alpha = 0$  & $\alpha = 1$    \\     
    \hline
    
    $h_\text{X}$     &  $R_g$          & $35.3^{+9.6}_{-2.8}$     & $39.7^{+9.3}_{-7.6}$     \\
    $\dot{m}$        &  $\dot{m}_\text{Edd}$         & $<0.010$                 & $0.04^{+0.023}_{-0.019}$      \\
    $E(B-V)$         &           & $0.09 \pm 0.01$           &  $0.08 \pm 0.02$    \\
    
    $\nu_{b,X}$      &   day$^{-1}$         & $0.052^{+0.019}_{-0.018}$    & $0.050^{+0.020}_{-0.022}$     \\
    $s_X $           &           & $2.76 \pm 0.25$       & $2.70^{+0.28}_{-0.27}$     \\
    $A_X$            &   day        & $0.037^{+0.018}_{-0.008}$    & $0.037^{+0.021}_{-0.012}$     \\    
    $\chi^2/df$     &            & 95.9/75                 & 95.2/75     \\
    $P_\text{null}$ &            & 5.2 \%                  & 5.8 \%     \\
    \tableline 
    \end{tabular}
    \label{tab:bestfit_storm}
\end{table}


\subsection{A multi-epoch analysis}
\label{sec:multi_epoch}

In \palio, we studied the HX, W2 and W1 PSDs using light curves from two different periods (56384-56548 and 56613-56876 MJD; periods P1 and P2\footnote{The analysis of Sect. \ref{sec:fit_psd_lags} considered the data of period P2.}, respectively). We concluded that the X-ray PSDs were consistent between the two periods, despite a lower X-ray flux state during P1. Although the UV flux was similarly lower during P1, the W2 and W1 PSDs were significantly different at this earlier period. We then suggested that a difference in the height of the X-ray corona could account for the observed trend. Here, we proceed to test this hypothesis.

To this end, we consider the data sets fitted in the previous section (i.e. the  \uvot, HX PSDs and the lag spectrum) and the HX, W2, and W1 PSDs of period P1, obtained from \palio. The two X-ray PSDs are fit by the same model of eq. \ref{eq:psd_xraymodel}, with the model parameters being tied for the two sets, except for the experimental noise. All the \uvot\ PSDs are modelled using eqs. \ref{eq:psd_conv} and \ref{eq:transfer_benpl_main}, while the lag spectrum is fit by the same model as in Sect. \ref{sec:fit_psd_lags} and its parameters were linked to the PSD model parameters of period P2. The black hole mass is kept fixed for both periods to $M_8=0.52$, while for the P1 power spectra we assumed $\Gamma_\text{X}=1.61$, which was found by \cite{2015A&A...575A..22M} when analyzing the broadband X-ray spectrum of the source using data that largely overlap with P1. As the X-ray flux during P1 was found to be on average a factor of 0.75 smaller than during P2, we further assumed $L_\text{X, P1} = 0.75 \cdot L_\text{X, P2} = 0.0026~L_\text{Edd}$. The accretion rate and the extinction were linked to be the same in the two periods with their value being left free to be determined by the fit. Finally, the corona height was a free parameter during the fit. As previously, the fit is performed twice, for both spin values.

\begin{figure}[]
\centering
\includegraphics[width=0.9\linewidth,height=\linewidth, trim={0 0 0 0}, clip]{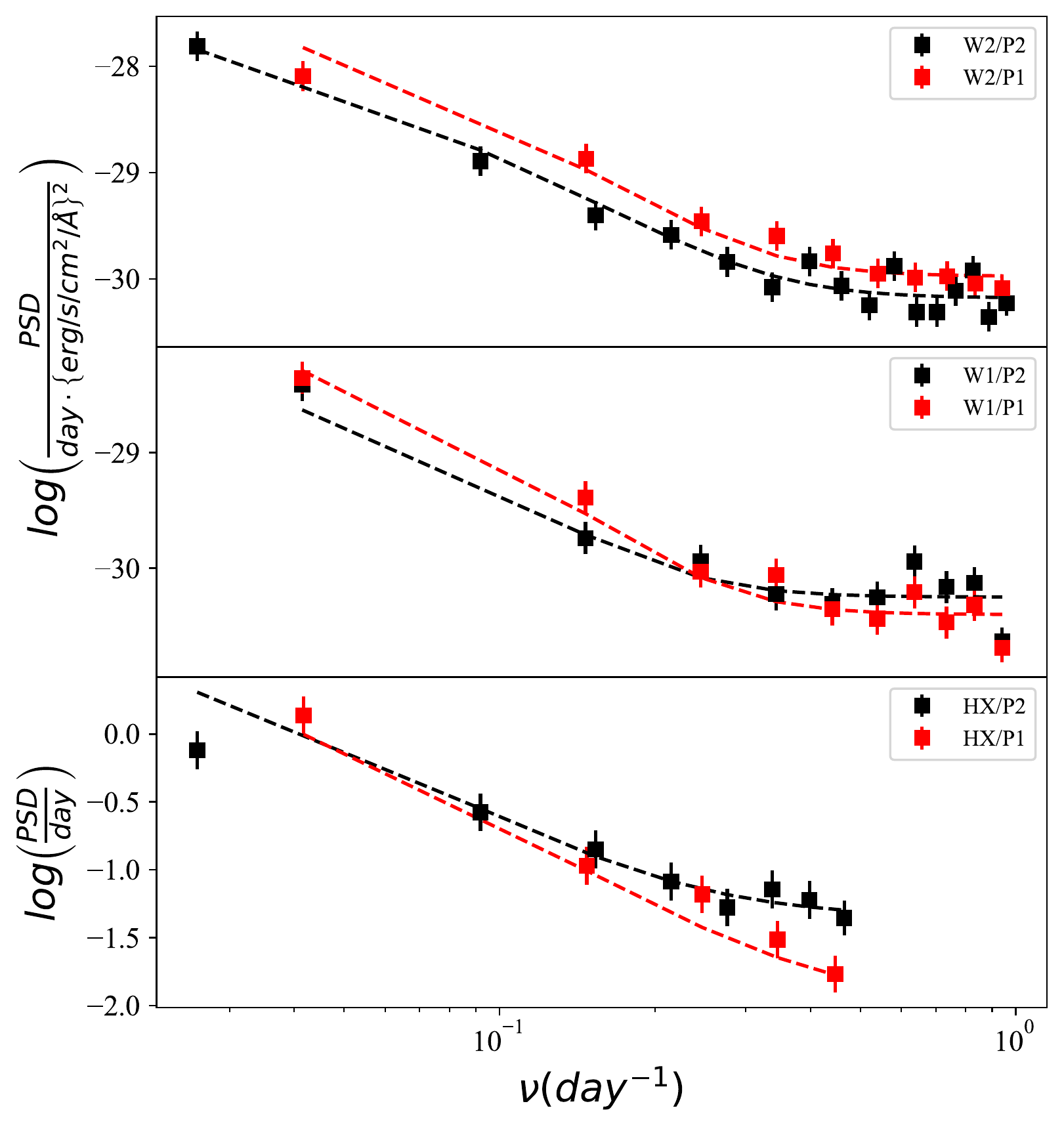}
\caption{W2 (top), W1 (middle), and HX (bottom panel) PSDs during both period P1 and P2 (red and black squares, respectively). The dashed lines (color coded in the same way) denote the best-fit model for $\alpha=1$, when only the corona height, $h_\text{X}$, is allowed to vary between the two periods. } 
\label{fig:w1w2psd_bestfit_2p}
\end{figure}

Table \ref{tab:bestfit_2p} lists the best-fit results. The corona height is indeed found to be significantly larger during P1, while the rest parameters are consistent within the errors with the values obtained when only the P2 data were considered. Figure \ref{fig:w1w2psd_bestfit_2p} shows the HX, W2, and W1 PSDs for both periods P1 and P2, together with the best-fit model for $\alpha=1$. The model fits the data well with no obvious systematic disagreement.

\begin{table}[]
    \centering
    \caption{Best fit results when we add the HX, W2, and W1 PSDs of period P1 to the analysis.}
    \begin{tabular}{llll}
    \hline
    Parameter        &   Units     &  $\alpha = 0$  & $\alpha = 1$    \\     
    \hline
    
    $h_\text{X}$/P2   &  $R_g$          & $26.4^{+7.0}_{-4.5}$      & $29.9^{+5.2}_{-5.1}$     \\
    $h_\text{X}$/P1   &  $R_g$          & $>63.2$     & $>73.6$     \\
    $\dot{m}$          &  $\dot{m}_\text{Edd}$         & $<0.012$                   & $0.063^{+0.023}_{-0.020}$      \\
    $E(B-V)$           &           & $0.11 \pm 0.01$            &  $0.10 \pm 0.01$    \\
    
    $\nu_{b,X}$        &   day$^{-1}$         & $0.055^{+0.024}_{-0.025}$          & $0.054^{+0.025}_{-0.027}$     \\
    $s_X $             &          & $2.50^{+0.26}_{-0.25}$       & $2.45^{+0.27}_{-0.25}$     \\
    $A_X$              &   day        & $0.069^{+0.038}_{-0.018}$    & $0.070^{+0.041}_{-0.018}$     \\
    
    $\chi^2/df$        &           & 131.1/97                 & 132.0/97     \\
    $P_\text{null}$    &           & 1.2\%                    & 1.0\%     \\
    \tableline 
    \end{tabular}
    \label{tab:bestfit_2p}
\end{table}


We also examined whether the fit is improved when more parameters are left to vary between P1 and P2. Firstly, we allowed the accretion rate to be different between the two epochs and repeated the fit. The values of accretion rate at the two periods were found to be consistent within the errors, while the improvement of the fit was not significant ($\Delta \chi^2/df =0.1/1$ and $5.5/1$ for $\alpha=0$ and $1$, respectively). We also checked whether the HX PSDs vary between the two epochs. We therefore repeated the fit by keeping the model parameters untied, but the fit did not improve significantly ($\Delta \chi^2/df = 12.5/3$ and $9.6/3$ for $\alpha=0$ and $1$, respectively).

Finally, we let the inclination angle free to be determined during the fit, while its value was linked between the two periods. Although the improvement of the fit was not statistically significant, it is worth noting that the new fit resulted in a best-fit value of $\theta = 53.1^{+7.7}_{-8.3} ~\text{degrees}$, which is fully consistent with the value of $\theta = 55.9^{+12.1}_{-6.9} ~\text{degrees}$ found by \cite{2021arXiv211001249D}, who studied the average spectral energy distribution of NGC 5548. 

The above analysis demonstrates how the PSD study can be used to investigate the dynamic evolution of AGN. The interpretation of the variable height is further discussed in Sect. \ref{sec:dynamic_evol}.

\section{Discussion} 
\label{sec:discuss}

\subsection{The \uvot\ PSD shape}
\label{sec:uvotpsd}

If X-rays illuminate the disk in AGN, then the \uvot\ PSDs are uniquely determined by the X-ray PSD, as shown in eq. \ref{eq:psd_conv}. We studied in detail the shape of the disk transfer function in AGN over a wide range of physical parameters. The results allow us to infer the broad characteristics of the AGN PSD in the \uvot\ band, if we consider a typical X-ray PSD shape.

For instance, let us assume a typical X-ray PSD following a bending power law with a low-frequency slope of -1, a bending frequency $\nu_{b,X}$, and a high-frequency slope of $s_{X} \simeq -2$, or even smaller, which has been found to describe well the X-ray power spectrum of numerous Seyfert galaxies \citep[e.g.][]{2003ApJ...593...96M, 2004MNRAS.348..783M, 2012A&A...544A..80G}. We can then predict the general shape of the \uvot\ PSDs in AGN, if the variability in these wavebands is due to X-ray reprocessing. 

The transfer function of Sect. \ref{sec:predict} shows that the \uvot\ PSD will feature the same shape as the X-ray PSD at low frequencies, whereas at high frequencies the \uvot\ PSD shall be dropping significantly faster than the X-ray one. In fact, the transfer function predicts the existence of a second bending frequency, $\nu_b$, in the \uvot\ PSD (Fig. \ref{fig:transfer_fidvals}), that is in addition to the X-ray PSD bending frequency $\nu_{b,X}$, which will also be imprinted in the \uvot\ PSD. Under the assumption of X-ray disk illumination, the \uvot\ PSD has one more bending frequency than the X-ray PSD. It should be stressed though that the detection of these bends is always subject to data quality and it seems challenging to probe the presence of multiple bendings using current data sets.

Furthermore, it would be interesting to compare $\nu_{b}$ to the X-ray PSD bend, $\nu_{b,X}$, and infer whether one is sytematically smaller than the other. However, such a comparison is not straightforward. The X-ray PSD bend, the physical origin of which is still unclear, scales with the black hole mass and, most likely, the source luminosity and spans values from around $10^{-7}$ Hz to $10^{-3}$ Hz \citep[e.g.][]{2003ApJ...593...96M, 2006Natur.444..730M, 2012A&A...544A..80G}. On the contrary, $\nu_{b}$, which appears due to the finite time needed for X-ray variations to illuminate the entire disk, depends on several physical properties (Appendix \ref{sec:a1}) and our analysis indicates that its value ranges from $10^{-7}$ to $10^{-5}$ Hz for the considered wavebands and parameter range (Figs. \ref{fig:transfer_ratio_mass}-\ref{fig:transfer_ratio_incl}). Therefore, depending on their physical properties, some sources will feature $\nu_{b,X} > \nu_{b}$, while for other sources $\nu_{b,X} < \nu_{b}$ will hold. In any case, given the generic shape of the X-ray PSD assumed above, the \uvot\ PSD is predicted to be the following.

If $\nu_{b,X} < \nu_{b}$, the \uvot\ PSD will feature the same shape as the X-ray PSD, scaled to the appropriate amplitude, up to the frequency $\nu_{b}$. In higher frequencies, the \uvot\ PSD follows a steeper power-law decline with a slope equal to the X-ray PSD slope plus the slope $s$ of eq. \ref{eq:transfer_benpl_main}. On the other hand, if $\nu_{b,X} > \nu_{b}$, the \uvot\ PSD shall follow a power law with a slope -1 up to frequency $\nu_{b}$, after which the PSD will feature a steeper drop, with a slope of $-1-s$. At $\nu_{b,X}$, the \uvot\ PSD will feature a second bend and it will decrease even faster at higher frequencies. In general, the \uvot\ PSD decline with frequency will be at least as steep as that of the X-ray PSD for a given source. This prediction can be tested against previous observational results.

\subsection{Comparing model predictions to observational data}
\label{sec:model_vs_obs}

Studying the X-ray PSD of a small sample of AGN, \cite{2003ApJ...593...96M} found that it can be reproduced by a power law wih typical slopes between -1 and -2. \cite{2018ApJ...857..141S} analyzed the \textit{Kepler} light curves of an AGN sample and deduced that the optical PSDs, which probe a similar frequency range to the X-ray PSDs considered by \cite{2003ApJ...593...96M}, feature a power law shape with typical slopes between -2 and -3. The fact that the \uvot\ PSD slopes are on average steeper than those in the X-ray PSD is consistent with our model predictions and suggests that the corresponding $\nu_{b}$ of the considered AGN lie within the probed frequency range of the above study or at even lower frequencies.

Perhaps, the best studied optical PSD of an AGN is that of Zw 229-15. This source is the brightest and longest monitored AGN in the \textit{Kepler} sample. \cite{2014ApJ...795....2E} showed that the \textit{Kepler} PSD of Zw 229-15 follows a bending power law, with a slope of -2 at low frequencies and of -4.5 at high frequencies, whereas the bending frequency was estimated to be $f_b = 0.18 ~\text{day}^{-1}$. While the X-ray PSD of this source is undetermined, assuming that it features a power law shape with a slope of -2 in the probed frequencies, the \textit{Kepler} PSD supports the scenario of an X-ray illuminated disk. Assuming a black hole mass of $M_\text{BH} = 10^7~ M_\odot$ and an accretion rate of $\dot{m} = 0.05 ~\dot{m}_\text{Edd}$ \citep{2011ApJ...732..121B}, and typical values for $\Gamma_\text{X} = 2$, $L_\text{X} = 0.01 ~L_\text{Edd}$, $\theta = 40 ^\circ$, and $h_\text{X} = 10 ~R_g$, eq. \ref{eq:transfer_benpl} predicts a bending frequency at 0.14 (0.21) day$^{-1}$ and a high-frequency slope of 4.2 (4.2), for a black hole spin $\alpha = 0$ (1). These values are in remarkable agreement with the values estimated by \cite{2014ApJ...795....2E} within the errors. 

Further, these authors noticed an excess variability over a power law decline in the PSD of Zw 229-15 at frequencies higher than around $10^{-5}$ Hz. Although this excess is probably due to unaccounted systematics of the \textit{Kepler} light curve \citep[see extensive discussion in][]{2014ApJ...795....2E}, we wish to highlight that the X-ray disk reprocessing results in similar variability excess (or, potentially, deficits) on top of a power-law decline at high frequencies (see, for example, the g' band panel in Fig. \ref{fig:transfer_ratio_height}). Therefore, it is interesting to note that our model can explain the shape of the observed PSD and can also account for the detected high-frequency variability excess, if real.

Finally, our model predicts that emission at longer wavelengths will be less variable than that of shorter wavelengths (Figs. \ref{fig:transfer_ratio_mass}-\ref{fig:transfer_ratio_incl}), as the former originate from a more extended region. Such an anticorrelation between the variability amplitude and the wavelength has been observed in a number of studies \citep[e.g.][]{2010ApJ...721.1014M, 2016A&A...585A.129S}, which seems to further support the scenario of an X-ray illuminated disk.

\subsection{PSD modelling of NGC 5548}

We applied our model to the case of NGC 5548, which is one of a few sources with contemporaneous, long and dense enough X-ray and \uvot\ light curves that allow such an analysis to be performed. We demonstrated that disk reprocessing of X-rays explains well the observed PSDs in the different wavebands as well as the interband time lags (Fig. \ref{fig:psdlags_bestfit}), while constraints for the physical properties of this source were obtained (Table \ref{tab:bestfit_storm}). To the best of our knowledge, this is the first time that a physical model is able to reproduce both the \uvot\ lags and the power spectra of an AGN.

Furthermore, \cite{2021arXiv211001249D} found that the broadband spectral energy distribution of this source can also be well reproduced assuming an X-ray illuminated accretion disk. We may, therefore, conclude that the thermal reprocessing of X-rays by the disk is the most favorable physical scenario for driving the \uvot\ variability at the considered time scales, at least in the case of NGC 5548.

We plan to extend our analysis to more sources in order to further test the X-ray reprocessing scenario. We also intend to update our model so that the energy spectrum, the lag spectrum, and the power spectra of a source can be simultaneously fit in a self-consistent way.


\subsection{Dynamic evolution of the X-ray source}
\label{sec:dynamic_evol}

Analysing the UV light curve of NGC 5548 in two different periods, we found that the normalizations of the UV PSDs vary in time (see also \palio). We fitted the PSDs in both epochs and we inferred that the PSD variations can be explained by the dynamic evolution of the X-ray source. More precisely, the height of the X-ray corona was found to vary by more than a factor of 2 within a few months, while the remaining parameters (including the X-ray PSD) were deduced to remain the same.

A similar evolution was recently reported by \cite{2020MNRAS.498.3184C}, when considering the X-ray reverberation lags and energy spectrum of a highly variable AGN \citep[see also][]{2020NatAs...4..597A}. The X-ray corona of that source was found to feature a variable height on time scales of days, suggesting that the geometry of the X-ray source may vary on both short and long time scales.

Moreover, these authors showed that the corona height increases with increasing X-ray luminosity. This is in contrast to our results, as we estimated a higher $h_\text{X}$ for the period P1, during which the observed X-ray flux is significantly smaller. Perharps, the X-ray corona features a different evolution pattern on long and short time scales. It would be interesting to further study this point once variations of the corona height in more sources are detected.

However, it should be mentioned that the dynamic evolution of the X-ray source on short time scales may challenge the rigorousness of the PSD fitting analysis, as this depends on the assumption of a steady system (Sect. \ref{sec:bkg}). A detailed investigation of how a variable $h_\text{X}$ affects our analysis and of any potential biases to our results is not easy because this depends strongly on the unknown correlation between the corona height and the other physical parameters as well as on the unknown time scale and amplitude of the $h_\text{X}$ variations. It is worth noticing, though, that the time lag spectrum, the energy spectrum, and the PSDs are all well reproduced by similar values of the physical parameters. The $h_\text{X}$ variability, if any, seems to affect all of them in a similar way and the resulted best-fit values shall, therefore, be representative of the average physical properties.

\section{Conclusions} 
\label{sec:conclude}

We conducted a detailed investigation of the expected disk variability in the case of X-ray illumination of the accretion disk and we studied its dependence on the different physical properties. Further, we developed an analytical description of the transfer function as a function of these physical properties, which we used to model simultaneously the observed PSDs and time lag spectrum of NGC 5548. Our main findings can be summarised as follows:

\begin{enumerate}
  \item Assuming the X-ray illumination of the disk results in strong predictions for the \uvot\ PSDs, such as the existence of a bending frequency at a specific range, the steep PSD slope at high frequencies, and the anti-correlation between the variability amplitude and wavelength, which can be tested against observations.
  
  \item In the case of NGC 5548, the observed PSDs and time lags are all well reproduced by our model, implying that the disk reprocessing of X-rays can account for the full variability properties of the \uvot\ emission of this source.
  
  \item The UV PSDs of NGC 5548 were found to vary between two periods separated by a few months. We inferred that this PSD variation is well explained if the height of the X-ray source varies between the two periods. Combining with recent observational results, our results indicate that the X-ray source of AGN is dynamic, changing its position (and potentially, its spatial extention) on various time scales.
  
\end{enumerate}

\section*{Acknowledgments}

CP acknowledges financial support from the Swiss National Science Foundation (SNF). ESK acknowledges financial support from the Centre National d’Etudes Spatiales (CNES). MD acknowledges the Czech MEYS grant LTAUSA17095 that supports international collaboration in relativistic astrophysics. MD acknowledges the Czech Science Foundation grant No. 21-06825X and the institutional support from RVO:67985815.

\bibliography{psd_predict}{}

\begin{thebibliography}{}
\expandafter\ifx\csname natexlab\endcsname\relax\def\natexlab#1{#1}\fi
\providecommand{\url}[1]{\href{#1}{#1}}
\providecommand{\dodoi}[1]{doi:~\href{http://doi.org/#1}{\nolinkurl{#1}}}
\providecommand{\doeprint}[1]{\href{http://ascl.net/#1}{\nolinkurl{http://ascl.net/#1}}}
\providecommand{\doarXiv}[1]{\href{https://arxiv.org/abs/#1}{\nolinkurl{https://arxiv.org/abs/#1}}}

\bibitem[{{Alston} {et~al.}(2020){Alston}, {Fabian}, {Kara}, {Parker},
  {Dovciak}, {Pinto}, {Jiang}, {Middleton}, {Miniutti}, {Walton}, {Wilkins},
  {Buisson}, {Caballero-Garcia}, {Cackett}, {De Marco}, {Gallo}, {Lohfink},
  {Reynolds}, {Uttley}, {Young}, \& {Zogbhi}}]{2020NatAs...4..597A}
{Alston}, W.~N., {Fabian}, A.~C., {Kara}, E., {et~al.} 2020, Nature Astronomy,
  4, 597, \dodoi{10.1038/s41550-019-1002-x}

\bibitem[{{Ar{\'e}valo} \& {Uttley}(2006)}]{2006MNRAS.367..801A}
{Ar{\'e}valo}, P., \& {Uttley}, P. 2006, \mnras, 367, 801,
  \dodoi{10.1111/j.1365-2966.2006.09989.x}

\bibitem[{{Arnaud}(1996)}]{1996ASPC..101...17A}
{Arnaud}, K.~A. 1996, in Astronomical Society of the Pacific Conference Series,
  Vol. 101, Astronomical Data Analysis Software and Systems V, ed. G.~H.
  {Jacoby} \& J.~{Barnes}, 17

\bibitem[{{Barth} {et~al.}(2011){Barth}, {Nguyen}, {Malkan}, {Filippenko},
  {Li}, {Gorjian}, {Joner}, {Bennert}, {Botyanszki}, {Cenko}, {Childress},
  {Choi}, {Comerford}, {Cucciara}, {da Silva}, {Duch{\^e}ne}, {Fumagalli},
  {Ganeshalingam}, {Gates}, {Gerke}, {Griffith}, {Harris}, {Hintz}, {Hsiao},
  {Kandrashoff}, {Keel}, {Kirkman}, {Kleiser}, {Laney}, {Lee}, {Lopez}, {Lowe},
  {Moody}, {Morton}, {Nierenberg}, {Nugent}, {Pancoast}, {Rex}, {Rich},
  {Silverman}, {Smith}, {Sonnenfeld}, {Suzuki}, {Tytler}, {Walsh}, {Woo},
  {Yang}, \& {Zeisse}}]{2011ApJ...732..121B}
{Barth}, A.~J., {Nguyen}, M.~L., {Malkan}, M.~A., {et~al.} 2011, \apj, 732,
  121, \dodoi{10.1088/0004-637X/732/2/121}

\bibitem[{{Bentz} \& {Katz}(2015)}]{2015PASP..127...67B}
{Bentz}, M.~C., \& {Katz}, S. 2015, \pasp, 127, 67, \dodoi{10.1086/679601}

\bibitem[{{Caballero-Garc{\'\i}a} {et~al.}(2018){Caballero-Garc{\'\i}a},
  {Papadakis}, {Dov{\v{c}}iak}, {Bursa}, {Epitropakis}, {Karas}, \&
  {Svoboda}}]{2018MNRAS.480.2650C}
{Caballero-Garc{\'\i}a}, M.~D., {Papadakis}, I.~E., {Dov{\v{c}}iak}, M.,
  {et~al.} 2018, \mnras, 480, 2650, \dodoi{10.1093/mnras/sty1990}

\bibitem[{{Caballero-Garc{\'\i}a} {et~al.}(2020){Caballero-Garc{\'\i}a},
  {Papadakis}, {Dov{\v{c}}iak}, {Bursa}, {Svoboda}, \&
  {Karas}}]{2020MNRAS.498.3184C}
---. 2020, \mnras, 498, 3184, \dodoi{10.1093/mnras/staa2554}

\bibitem[{{Cackett} {et~al.}(2021){Cackett}, {Bentz}, \&
  {Kara}}]{2021iSci...24j2557C}
{Cackett}, E.~M., {Bentz}, M.~C., \& {Kara}, E. 2021, iScience, 24, 102557,
  \dodoi{10.1016/j.isci.2021.102557}

\bibitem[{{Cardelli} {et~al.}(1989){Cardelli}, {Clayton}, \&
  {Mathis}}]{1989ApJ...345..245C}
{Cardelli}, J.~A., {Clayton}, G.~C., \& {Mathis}, J.~S. 1989, \apj, 345, 245,
  \dodoi{10.1086/167900}

\bibitem[{{Clavel} {et~al.}(1992){Clavel}, {Nandra}, {Makino}, {Pounds},
  {Reichert}, {Urry}, {Wamsteker}, {Peracaula-Bosch}, {Stewart}, \&
  {Otani}}]{1992ApJ...393..113C}
{Clavel}, J., {Nandra}, K., {Makino}, F., {et~al.} 1992, \apj, 393, 113,
  \dodoi{10.1086/171490}

\bibitem[{{De Rosa} {et~al.}(2015){De Rosa}, {Peterson}, {Ely}, {Kriss},
  {Crenshaw}, {Horne}, {Korista}, {Netzer}, {Pogge}, {Ar{\'e}valo}, {Barth},
  {Bentz}, {Brandt}, {Breeveld}, {Brewer}, {Dalla Bont{\`a}}, {De
  Lorenzo-C{\'a}ceres}, {Denney}, {Dietrich}, {Edelson}, {Evans}, {Fausnaugh},
  {Gehrels}, {Gelbord}, {Goad}, {Grier}, {Grupe}, {Hall}, {Kaastra}, {Kelly},
  {Kennea}, {Kochanek}, {Lira}, {Mathur}, {McHardy}, {Nousek}, {Pancoast},
  {Papadakis}, {Pei}, {Schimoia}, {Siegel}, {Starkey}, {Treu}, {Uttley},
  {Vaughan}, {Vestergaard}, {Villforth}, {Yan}, {Young}, \&
  {Zu}}]{2015ApJ...806..128D}
{De Rosa}, G., {Peterson}, B.~M., {Ely}, J., {et~al.} 2015, \apj, 806, 128,
  \dodoi{10.1088/0004-637X/806/1/128}

\bibitem[{{Dovciak} {et~al.}(2021){Dovciak}, {Papadakis}, {Kammoun}, \&
  {Zhang}}]{2021arXiv211001249D}
{Dovciak}, M., {Papadakis}, I.~E., {Kammoun}, E.~S., \& {Zhang}, W. 2021, arXiv
  e-prints, arXiv:2110.01249.
\newblock \doarXiv{2110.01249}

\bibitem[{{Edelson} {et~al.}(2014){Edelson}, {Vaughan}, {Malkan}, {Kelly},
  {Smith}, {Boyd}, \& {Mushotzky}}]{2014ApJ...795....2E}
{Edelson}, R., {Vaughan}, S., {Malkan}, M., {et~al.} 2014, \apj, 795, 2,
  \dodoi{10.1088/0004-637X/795/1/2}

\bibitem[{{Edelson} {et~al.}(2015){Edelson}, {Gelbord}, {Horne}, {McHardy},
  {Peterson}, {Ar{\'e}valo}, {Breeveld}, {De Rosa}, {Evans}, {Goad}, {Kriss},
  {Brandt}, {Gehrels}, {Grupe}, {Kennea}, {Kochanek}, {Nousek}, {Papadakis},
  {Siegel}, {Starkey}, {Uttley}, {Vaughan}, {Young}, {Barth}, {Bentz},
  {Brewer}, {Crenshaw}, {Dalla Bont{\`a}}, {De Lorenzo-C{\'a}ceres}, {Denney},
  {Dietrich}, {Ely}, {Fausnaugh}, {Grier}, {Hall}, {Kaastra}, {Kelly},
  {Korista}, {Lira}, {Mathur}, {Netzer}, {Pancoast}, {Pei}, {Pogge},
  {Schimoia}, {Treu}, {Vestergaard}, {Villforth}, {Yan}, \&
  {Zu}}]{2015ApJ...806..129E}
{Edelson}, R., {Gelbord}, J.~M., {Horne}, K., {et~al.} 2015, \apj, 806, 129,
  \dodoi{10.1088/0004-637X/806/1/129}

\bibitem[{{Edelson} {et~al.}(2019){Edelson}, {Gelbord}, {Cackett}, {Peterson},
  {Horne}, {Barth}, {Starkey}, {Bentz}, {Brandt}, {Goad}, {Joner}, {Korista},
  {Netzer}, {Page}, {Uttley}, {Vaughan}, {Breeveld}, {Cenko}, {Done}, {Evans},
  {Fausnaugh}, {Ferland}, {Gonzalez-Buitrago}, {Gropp}, {Grupe}, {Kaastra},
  {Kennea}, {Kriss}, {Mathur}, {Mehdipour}, {Mudd}, {Nousek}, {Schmidt},
  {Vestergaard}, \& {Villforth}}]{2019ApJ...870..123E}
{Edelson}, R., {Gelbord}, J., {Cackett}, E., {et~al.} 2019, \apj, 870, 123,
  \dodoi{10.3847/1538-4357/aaf3b4}

\bibitem[{{Elvis} {et~al.}(1978){Elvis}, {Maccacaro}, {Wilson}, {Ward},
  {Penston}, {Fosbury}, \& {Perola}}]{1978MNRAS.183..129E}
{Elvis}, M., {Maccacaro}, T., {Wilson}, A.~S., {et~al.} 1978, \mnras, 183, 129,
  \dodoi{10.1093/mnras/183.2.129}

\bibitem[{{Epitropakis} {et~al.}(2016){Epitropakis}, {Papadakis},
  {Dov{\v{c}}iak}, {Pech{\'a}{\v{c}}ek}, {Emmanoulopoulos}, {Karas}, \&
  {McHardy}}]{2016A&A...594A..71E}
{Epitropakis}, A., {Papadakis}, I.~E., {Dov{\v{c}}iak}, M., {et~al.} 2016,
  \aap, 594, A71, \dodoi{10.1051/0004-6361/201527748}

\bibitem[{{Fausnaugh} {et~al.}(2016){Fausnaugh}, {Denney}, {Barth}, {Bentz},
  {Bottorff}, {Carini}, {Croxall}, {De Rosa}, {Goad}, {Horne}, {Joner},
  {Kaspi}, {Kim}, {Klimanov}, {Kochanek}, {Leonard}, {Netzer}, {Peterson},
  {Schn{\"u}lle}, {Sergeev}, {Vestergaard}, {Zheng}, {Zu}, {Anderson},
  {Ar{\'e}valo}, {Bazhaw}, {Borman}, {Boroson}, {Brandt}, {Breeveld}, {Brewer},
  {Cackett}, {Crenshaw}, {Dalla Bont{\`a}}, {De Lorenzo-C{\'a}ceres},
  {Dietrich}, {Edelson}, {Efimova}, {Ely}, {Evans}, {Filippenko}, {Flatland},
  {Gehrels}, {Geier}, {Gelbord}, {Gonzalez}, {Gorjian}, {Grier}, {Grupe},
  {Hall}, {Hicks}, {Horenstein}, {Hutchison}, {Im}, {Jensen}, {Jones},
  {Kaastra}, {Kelly}, {Kennea}, {Kim}, {Korista}, {Kriss}, {Lee}, {Lira},
  {MacInnis}, {Manne-Nicholas}, {Mathur}, {McHardy}, {Montouri}, {Musso},
  {Nazarov}, {Norris}, {Nousek}, {Okhmat}, {Pancoast}, {Papadakis}, {Parks},
  {Pei}, {Pogge}, {Pott}, {Rafter}, {Rix}, {Saylor}, {Schimoia}, {Siegel},
  {Spencer}, {Starkey}, {Sung}, {Teems}, {Treu}, {Turner}, {Uttley},
  {Villforth}, {Weiss}, {Woo}, {Yan}, \& {Young}}]{2016ApJ...821...56F}
{Fausnaugh}, M.~M., {Denney}, K.~D., {Barth}, A.~J., {et~al.} 2016, \apj, 821,
  56, \dodoi{10.3847/0004-637X/821/1/56}

\bibitem[{{Gonz{\'a}lez-Mart{\'\i}n} \& {Vaughan}(2012)}]{2012A&A...544A..80G}
{Gonz{\'a}lez-Mart{\'\i}n}, O., \& {Vaughan}, S. 2012, \aap, 544, A80,
  \dodoi{10.1051/0004-6361/201219008}

\bibitem[{{Grier} {et~al.}(2013){Grier}, {Martini}, {Watson}, {Peterson},
  {Bentz}, {Dasyra}, {Dietrich}, {Ferrarese}, {Pogge}, \&
  {Zu}}]{2013ApJ...773...90G}
{Grier}, C.~J., {Martini}, P., {Watson}, L.~C., {et~al.} 2013, \apj, 773, 90,
  \dodoi{10.1088/0004-637X/773/2/90}

\bibitem[{{Haardt} \& {Maraschi}(1993)}]{1993ApJ...413..507H}
{Haardt}, F., \& {Maraschi}, L. 1993, \apj, 413, 507, \dodoi{10.1086/173020}

\bibitem[{{Kammoun} {et~al.}(2021{\natexlab{a}}){Kammoun}, {Dov{\v{c}}iak},
  {Papadakis}, {Caballero-Garc{\'\i}a}, \& {Karas}}]{2021ApJ...907...20K}
{Kammoun}, E.~S., {Dov{\v{c}}iak}, M., {Papadakis}, I.~E.,
  {Caballero-Garc{\'\i}a}, M.~D., \& {Karas}, V. 2021{\natexlab{a}}, \apj, 907,
  20, \dodoi{10.3847/1538-4357/abcb93}

\bibitem[{{Kammoun} {et~al.}(2019){Kammoun}, {Papadakis}, \&
  {Dov{\v{c}}iak}}]{2019ApJ...879L..24K}
{Kammoun}, E.~S., {Papadakis}, I.~E., \& {Dov{\v{c}}iak}, M. 2019, \apjl, 879,
  L24, \dodoi{10.3847/2041-8213/ab2a72}

\bibitem[{{Kammoun} {et~al.}(2021{\natexlab{b}}){Kammoun}, {Papadakis}, \&
  {Dov{\v{c}}iak}}]{2021MNRAS.503.4163K}
---. 2021{\natexlab{b}}, \mnras, 503, 4163, \dodoi{10.1093/mnras/stab725}

\bibitem[{{Kara} {et~al.}(2015){Kara}, {Zoghbi}, {Marinucci}, {Walton},
  {Fabian}, {Risaliti}, {Boggs}, {Christensen}, {Fuerst}, {Hailey}, {Harrison},
  {Matt}, {Parker}, {Reynolds}, {Stern}, \& {Zhang}}]{2015MNRAS.446..737K}
{Kara}, E., {Zoghbi}, A., {Marinucci}, A., {et~al.} 2015, \mnras, 446, 737,
  \dodoi{10.1093/mnras/stu2136}

\bibitem[{{Kelly} {et~al.}(2014){Kelly}, {Becker}, {Sobolewska},
  {Siemiginowska}, \& {Uttley}}]{2014ApJ...788...33K}
{Kelly}, B.~C., {Becker}, A.~C., {Sobolewska}, M., {Siemiginowska}, A., \&
  {Uttley}, P. 2014, \apj, 788, 33, \dodoi{10.1088/0004-637X/788/1/33}

\bibitem[{{MacLeod} {et~al.}(2010){MacLeod}, {Ivezi{\'c}}, {Kochanek},
  {Koz{\l}owski}, {Kelly}, {Bullock}, {Kimball}, {Sesar}, {Westman}, {Brooks},
  {Gibson}, {Becker}, \& {de Vries}}]{2010ApJ...721.1014M}
{MacLeod}, C.~L., {Ivezi{\'c}}, {\v{Z}}., {Kochanek}, C.~S., {et~al.} 2010,
  \apj, 721, 1014, \dodoi{10.1088/0004-637X/721/2/1014}

\bibitem[{{Markowitz} {et~al.}(2003){Markowitz}, {Edelson}, {Vaughan},
  {Uttley}, {George}, {Griffiths}, {Kaspi}, {Lawrence}, {McHardy}, {Nandra},
  {Pounds}, {Reeves}, {Schurch}, \& {Warwick}}]{2003ApJ...593...96M}
{Markowitz}, A., {Edelson}, R., {Vaughan}, S., {et~al.} 2003, \apj, 593, 96,
  \dodoi{10.1086/375330}

\bibitem[{{Mastroserio} {et~al.}(2020){Mastroserio}, {Ingram}, \& {van der
  Klis}}]{2020MNRAS.498.4971M}
{Mastroserio}, G., {Ingram}, A., \& {van der Klis}, M. 2020, \mnras, 498, 4971,
  \dodoi{10.1093/mnras/staa2735}

\bibitem[{{Mathur} {et~al.}(2017){Mathur}, {Gupta}, {Page}, {Pogge},
  {Krongold}, {Goad}, {Adams}, {Anderson}, {Ar{\'e}valo}, {Barth}, {Bazhaw},
  {Beatty}, {Bentz}, {Bigley}, {Bisogni}, {Borman}, {Boroson}, {Bottorff},
  {Brandt}, {Breeveld}, {Brown}, {Brown}, {Cackett}, {Canalizo}, {Carini},
  {Clubb}, {Comerford}, {Coker}, {Corsini}, {Crenshaw}, {Croft}, {Croxall},
  {Dalla Bont{\`a}}, {Deason}, {Denney}, {De Lorenzo-C{\'a}ceres}, {De Rosa},
  {Dietrich}, {Edelson}, {Ely}, {Eracleous}, {Evans}, {Fausnaugh}, {Ferland},
  {Filippenko}, {Flatland}, {Fox}, {Gates}, {Gehrels}, {Geier}, {Gelbord},
  {Gorjian}, {Greene}, {Grier}, {Grupe}, {Hall}, {Henderson}, {Hicks},
  {Holmbeck}, {Holoien}, {Horenstein}, {Horne}, {Hutchison}, {Im}, {Jensen},
  {Johnson}, {Joner}, {Jones}, {Kaastra}, {Kaspi}, {Kelly}, {Kelly}, {Kennea},
  {Kim}, {Kim}, {Kim}, {King}, {Klimanov}, {Kochanek}, {Korista}, {Kriss},
  {Lau}, {Lee}, {Leonard}, {Li}, {Lira}, {Ma}, {MacInnis}, {Manne-Nicholas},
  {Malkan}, {Mauerhan}, {McGurk}, {McHardy}, {Montouri}, {Morelli}, {Mosquera},
  {Mudd}, {Muller-Sanchez}, {Musso}, {Nazarov}, {Netzer}, {Nguyen}, {Norris},
  {Nousek}, {Ochner}, {Okhmat}, {Ou-Yang}, {Pancoast}, {Papadakis}, {Parks},
  {Pei}, {Peterson}, {Pizzella}, {Poleski}, {Pott}, {Rafter}, {Rix}, {Runnoe},
  {Saylor}, {Schimoia}, {Schn{\"u}lle}, {Sergeev}, {Shappee}, {Shivvers},
  {Siegel}, {Simonian}, {Siviero}, {Skielboe}, {Somers}, {Spencer}, {Starkey},
  {Stevens}, {Sung}, {Tayar}, {Tejos}, {Turner}, {Uttley}, {Van Saders},
  {Vestergaard}, {Vican}, {Villanueva}, {Villforth}, {Weiss}, {Woo}, {Yan},
  {Young}, {Yuk}, {Zheng}, {Zhu}, \& {Zu}}]{2017ApJ...846...55M}
{Mathur}, S., {Gupta}, A., {Page}, K., {et~al.} 2017, \apj, 846, 55,
  \dodoi{10.3847/1538-4357/aa832b}

\bibitem[{{McHardy} {et~al.}(2006){McHardy}, {Koerding}, {Knigge}, {Uttley}, \&
  {Fender}}]{2006Natur.444..730M}
{McHardy}, I.~M., {Koerding}, E., {Knigge}, C., {Uttley}, P., \& {Fender},
  R.~P. 2006, \nat, 444, 730, \dodoi{10.1038/nature05389}

\bibitem[{{McHardy} {et~al.}(2004){McHardy}, {Papadakis}, {Uttley}, {Page}, \&
  {Mason}}]{2004MNRAS.348..783M}
{McHardy}, I.~M., {Papadakis}, I.~E., {Uttley}, P., {Page}, M.~J., \& {Mason},
  K.~O. 2004, \mnras, 348, 783, \dodoi{10.1111/j.1365-2966.2004.07376.x}

\bibitem[{{McHardy} {et~al.}(2018){McHardy}, {Connolly}, {Horne}, {Cackett},
  {Gelbord}, {Peterson}, {Pahari}, {Gehrels}, {Goad}, {Lira}, {Arevalo},
  {Baldi}, {Brandt}, {Breedt}, {Chand}, {Dewangan}, {Done}, {Elvis},
  {Emmanoulopoulos}, {Fausnaugh}, {Kaspi}, {Kochanek}, {Korista}, {Papadakis},
  {Rao}, {Uttley}, {Vestergaard}, \& {Ward}}]{2018MNRAS.480.2881M}
{McHardy}, I.~M., {Connolly}, S.~D., {Horne}, K., {et~al.} 2018, \mnras, 480,
  2881, \dodoi{10.1093/mnras/sty1983}

\bibitem[{{Mehdipour} {et~al.}(2015){Mehdipour}, {Kaastra}, {Kriss}, {Cappi},
  {Petrucci}, {Steenbrugge}, {Arav}, {Behar}, {Bianchi}, {Boissay},
  {Branduardi-Raymont}, {Costantini}, {Ebrero}, {Di Gesu}, {Harrison}, {Kaspi},
  {De Marco}, {Matt}, {Paltani}, {Peterson}, {Ponti}, {Pozo Nu{\~n}ez}, {De
  Rosa}, {Ursini}, {de Vries}, {Walton}, \& {Whewell}}]{2015A&A...575A..22M}
{Mehdipour}, M., {Kaastra}, J.~S., {Kriss}, G.~A., {et~al.} 2015, \aap, 575,
  A22, \dodoi{10.1051/0004-6361/201425373}

\bibitem[{{Mushotzky} {et~al.}(2011){Mushotzky}, {Edelson}, {Baumgartner}, \&
  {Gandhi}}]{2011ApJ...743L..12M}
{Mushotzky}, R.~F., {Edelson}, R., {Baumgartner}, W., \& {Gandhi}, P. 2011,
  \apjl, 743, L12, \dodoi{10.1088/2041-8205/743/1/L12}

\bibitem[{{Novikov} \& {Thorne}(1973)}]{1973blho.conf..343N}
{Novikov}, I.~D., \& {Thorne}, K.~S. 1973, in Black Holes (Les Astres Occlus),
  343--450

\bibitem[{{Panagiotou} {et~al.}(2020){Panagiotou}, {Papadakis}, {Kammoun}, \&
  {Dov{\v{c}}iak}}]{2020MNRAS.499.1998P}
{Panagiotou}, C., {Papadakis}, I.~E., {Kammoun}, E.~S., \& {Dov{\v{c}}iak}, M.
  2020, \mnras, 499, 1998, \dodoi{10.1093/mnras/staa2920}

\bibitem[{{Papadakis} {et~al.}(2016){Papadakis}, {Pech{\'a}{\v{c}}ek},
  {Dov{\v{c}}iak}, {Epitropakis}, {Emmanoulopoulos}, \&
  {Karas}}]{2016A&A...588A..13P}
{Papadakis}, I., {Pech{\'a}{\v{c}}ek}, T., {Dov{\v{c}}iak}, M., {et~al.} 2016,
  \aap, 588, A13, \dodoi{10.1051/0004-6361/201527246}

\bibitem[{{Papadakis} {et~al.}(2007){Papadakis}, {Ioannou}, \&
  {Kazanas}}]{2007ApJ...661...38P}
{Papadakis}, I.~E., {Ioannou}, Z., \& {Kazanas}, D. 2007, \apj, 661, 38,
  \dodoi{10.1086/513307}

\bibitem[{{Peterson} {et~al.}(2004){Peterson}, {Ferrarese}, {Gilbert}, {Kaspi},
  {Malkan}, {Maoz}, {Merritt}, {Netzer}, {Onken}, {Pogge}, {Vestergaard}, \&
  {Wandel}}]{2004ApJ...613..682P}
{Peterson}, B.~M., {Ferrarese}, L., {Gilbert}, K.~M., {et~al.} 2004, \apj, 613,
  682, \dodoi{10.1086/423269}

\bibitem[{Priestley(1981)}]{nla.cat-vn2888327}
Priestley, M.~B. 1981, Spectral analysis and time series / M.B. Priestley
  (Academic Press London ; New York), 2 v. (xvii, [45], 890 p.) :

\bibitem[{{Reis} \& {Miller}(2013)}]{2013ApJ...769L...7R}
{Reis}, R.~C., \& {Miller}, J.~M. 2013, \apjl, 769, L7,
  \dodoi{10.1088/2041-8205/769/1/L7}

\bibitem[{{Ross} {et~al.}(1992){Ross}, {Fabian}, \&
  {Mineshige}}]{1992MNRAS.258..189R}
{Ross}, R.~R., {Fabian}, A.~C., \& {Mineshige}, S. 1992, \mnras, 258, 189,
  \dodoi{10.1093/mnras/258.1.189}

\bibitem[{{Salpeter}(1964)}]{1964ApJ...140..796S}
{Salpeter}, E.~E. 1964, \apj, 140, 796, \dodoi{10.1086/147973}

\bibitem[{{Schlafly} \& {Finkbeiner}(2011)}]{2011ApJ...737..103S}
{Schlafly}, E.~F., \& {Finkbeiner}, D.~P. 2011, \apj, 737, 103,
  \dodoi{10.1088/0004-637X/737/2/103}

\bibitem[{{Shakura} \& {Sunyaev}(1973)}]{1973A&A....24..337S}
{Shakura}, N.~I., \& {Sunyaev}, R.~A. 1973, \aap, 500, 33

\bibitem[{{Shields}(1999)}]{1999PASP..111..661S}
{Shields}, G.~A. 1999, \pasp, 111, 661, \dodoi{10.1086/316378}

\bibitem[{{Simm} {et~al.}(2016){Simm}, {Salvato}, {Saglia}, {Ponti},
  {Lanzuisi}, {Trakhtenbrot}, {Nandra}, \& {Bender}}]{2016A&A...585A.129S}
{Simm}, T., {Salvato}, M., {Saglia}, R., {et~al.} 2016, \aap, 585, A129,
  \dodoi{10.1051/0004-6361/201527353}

\bibitem[{{Smith} {et~al.}(2018){Smith}, {Mushotzky}, {Boyd}, {Malkan},
  {Howell}, \& {Gelino}}]{2018ApJ...857..141S}
{Smith}, K.~L., {Mushotzky}, R.~F., {Boyd}, P.~T., {et~al.} 2018, \apj, 857,
  141, \dodoi{10.3847/1538-4357/aab88d}

\bibitem[{{Sun} {et~al.}(2020){Sun}, {Xue}, {Brandt}, {Gu}, {Trump}, {Cai},
  {He}, {Lin}, {Liu}, \& {Wang}}]{2020ApJ...891..178S}
{Sun}, M., {Xue}, Y., {Brandt}, W.~N., {et~al.} 2020, \apj, 891, 178,
  \dodoi{10.3847/1538-4357/ab789e}

\bibitem[{{Uttley} {et~al.}(2003){Uttley}, {Edelson}, {McHardy}, {Peterson}, \&
  {Markowitz}}]{2003ApJ...584L..53U}
{Uttley}, P., {Edelson}, R., {McHardy}, I.~M., {Peterson}, B.~M., \&
  {Markowitz}, A. 2003, \apjl, 584, L53, \dodoi{10.1086/373887}

\bibitem[{{Zoghbi} {et~al.}(2014){Zoghbi}, {Cackett}, {Reynolds}, {Kara},
  {Harrison}, {Fabian}, {Lohfink}, {Matt}, {Balokovic}, {Boggs}, {Christensen},
  {Craig}, {Hailey}, {Stern}, \& {Zhang}}]{2014ApJ...789...56Z}
{Zoghbi}, A., {Cackett}, E.~M., {Reynolds}, C., {et~al.} 2014, \apj, 789, 56,
  \dodoi{10.1088/0004-637X/789/1/56}

\end{thebibliography}
\bibliographystyle{aasjournal}

\appendix

\section{Parametric equations and their quality}
\label{sec:a1}

Plots of the the squared modulus of the transfer functions are shown in Appendix \ref{sec:plots}. As we already discussed, $|\Gamma_\lambda(\nu)|^2$ follows a power-like shape at high frequencies and then it flattens and reaches a constant value at low frequencies. We parametrized $|\Gamma_\lambda(\nu)|^2$ as a function of frequency with a bending power law model of the form:

\begin{equation}
\label{eq:transfer_benpl}
    |\Gamma_{approx,\lambda}(\nu)|^2(\vec{\mu})  = N(\lambda, \mu)  \cdot \frac{1}{1 +(\frac{\nu}{\nu_b(\lambda, \mu) })^{\large{s(\lambda, \mu) }}},
\end{equation}

\noindent where $\nu$ is the temporal frequency, $\lambda$ denotes the wavelength, and $\vec{\mu}$  represents the set of the model parameters, i.e. $\vec{\mu}=(M_8, \dot{m}, h_\text{cor}, L_X, \Gamma_X, \theta).$ These parameters determine the state of the disk/corona system.  The function defined by eq.\,(\ref{eq:transfer_benpl}) has a power-law like shape with a slope of $s$ at frequencies significantly higher than the bending frequency, $\nu_b$. At low frequencies ($\nu<<\nu_b$), the function is constant, i.e. $|\Gamma_{approx, \lambda}(\nu)|^2(\lambda, \mu)  \sim  N(\lambda, \mu)$. We provide below analytical functions that determine how the normalization, $N$,  break frequency, $\nu_b$, and the slope, $s$, depend on  $M_8, \dot{m}, h_\text{cor}, L_X, \Gamma_X, \theta$, and the wavelength $\lambda$.

The break frequency, measured in units of Hz, is given by:

\begin{equation}
    \nu_b(\lambda, \mu) = a_\nu (h_\text{X}) \cdot a_\nu(M_8) \cdot a_\nu (\dot{m}) \cdot a_\nu (L_X) \cdot a_\nu (\Gamma_X) \cdot a_\nu (\theta) \cdot \lambda_{1950}^{-[b_\nu (h_\text{X})+b_\nu(M_8) + b_\nu (\dot{m}) + b_\nu (L_X) + b_\nu (\Gamma_X) + b_\nu (\theta)]} \\
\end{equation}

\noindent where $\lambda_{1950} = \frac{\lambda}{1950 \AA}$, and for $\alpha=0$:

\begin{eqnarray*}
&a_\nu(h_\text{X}) &= 6.33 \cdot 10^{-6}  \cdot \frac{1}{1 +(\frac{h_\text{X}}{290.4})^{0.718}} \\
&a_\nu(M_8) &= 0.22 \cdot M_8^{-0.652} + 0.0129 \\
&a_\nu(\dot{m}) &= 2.251  \cdot \frac{1}{1 +(\frac{\dot{m}}{0.030})^{0.449}}\\
&a_\nu(L_X) &= 1.119  \cdot \frac{1}{1 +(\frac{L_X}{0.426})^{0.568}}\\
&a_\nu(\Gamma_X) &= -0.246 \cdot \Gamma_X^2 + 0.862 \cdot \Gamma_X + 0.260\\
&a_\nu(\theta) &= 1.082  \cdot \frac{1}{1 +(\frac{\theta}{96.3})^{2.84}}\\
\\
&b_\nu(h_\text{X}) &= -9.929 \cdot 10^{-4} \cdot h_\text{X} + 1.3\\
&b_\nu(M_8) &= -0.0564 \cdot M_8^2 + 0.1 \cdot M_8 - 0.0094\\
&b_\nu(\dot{m}) &= -0.098 \cdot \dot{m} + 0.0049 \\
&b_\nu(L_X) &= 1.197 \cdot L_X^{-0.0147} - 1.2808 \\
&b_\nu(\Gamma_X) &= -0.0514 \cdot \Gamma_X^2 + 0.1779 \cdot \Gamma_X - 0.1502\\
&b_\nu(\theta) &= 3.966 \cdot 10^{-6} \cdot \theta^3 - 3.363 \cdot 10^{-4} \cdot \theta^2 + 5.519 \cdot 10^{-3} \cdot \theta + 0.0635\\
\end{eqnarray*}

In the case of $\alpha=1$, the above relations are modified to:

\begin{eqnarray*}
&a_\nu(h_\text{X}) &= 1.421 \cdot 10^{-5}  \cdot \frac{1}{1 +(\frac{h_\text{X}}{49.91})^{0.549}} \\
&a_\nu(M_8) &= 0.2137 \cdot M_8^{-0.66} + 0.0231 \\
&a_\nu(\dot{m}) &= 1.304  \cdot \frac{1}{1 +(\frac{\dot{m}}{0.388})^{0.582}}\\
&a_\nu(L_X) &= 1.551  \cdot \frac{1}{1 +(\frac{L_X}{0.0376})^{0.45}}\\
&a_\nu(\Gamma_X) &= -0.4799 \cdot \Gamma_X^2 + 1.6658 \cdot \Gamma_X - 0.4122\\
&a_\nu(\theta) &= 1.114  \cdot \frac{1}{1 +(\frac{\theta}{110})^{2.144}}\\
\\
&b_\nu(h_\text{X}) &= -9.679 \cdot 10^{-4} \cdot h_\text{X} + 1.285\\
&b_\nu(M_8) &= 0.1383 \cdot M_8^2 - 0.184 \cdot M_8 + 0.017\\
&b_\nu(\dot{m}) &= -0.014 \cdot \dot{m} + 0.001 \\
&b_\nu(L_X) &= 1.176 \cdot L_X^{-0.0166} - 1.269 \\
&b_\nu(\Gamma_X) &= -0.0601 \cdot \Gamma_X^2 + 0.2077 \cdot \Gamma_X - 0.175\\
&b_\nu(\theta) &= 1.332 \cdot 10^{-6} \cdot \theta^3 - 5.349 \cdot 10^{-5} \cdot \theta^2 - 1.434 \cdot 10^{-3} \cdot \theta + 0.0577\\
\end{eqnarray*}

Similarly, the slope, $s$, is given by:

\begin{equation}
\label{eq:transfer_slope}
    s(\lambda, \mu) = a_s(h_\text{X}) \cdot a_s(M_8) \cdot a_s (\dot{m}) \cdot a_s (L_X) \cdot a_s (\Gamma_X) \cdot a_s (\theta) \cdot \lambda_{1950}^{-[b_s (h_\text{X}) + b_s (\dot{m}) + b_s (L_X) + b_s (\Gamma_X) + b_s (\theta]} \\
\end{equation}

where, for $\alpha=0$:

\begin{eqnarray*}
    &a_s(h_\text{X}) &= 1.8263 \cdot h_\text{X}^{0.0204} \\
    &a_s(M_8) &= 1 \\
    &a_s(\dot{m}) &= 1.056 \cdot \dot{m}^{0.0181} \\
    &a_s(L_X) &= 1.073 \cdot L_X^{0.0153} \\
    &a_s(\Gamma_X) &= 0.0506 \cdot \Gamma^2 - 0.1755 \cdot \Gamma + 1.1484\\
    &a_s(\theta) &= 0.7901 \cdot cos^2(\theta) - 0.2317 \cdot cos(\theta) + 0.7138\\
\\
    &b_s(h_\text{X}) &= -4.88 \cdot 10^{-6} \cdot h_\text{X}^2 + 9.443 \cdot 10^{-4} \cdot h_\text{X} - 0.133 \\
    &b_s(\dot{m}) &= - 0.0703 \cdot \dot{m}^2 + 0.059 \cdot \dot{m} - 0.0028 \\
    &b_s(L_X) &= -0.097 \cdot L_X^{-0.045} + 0.1193 \\
    &b_s(\Gamma_X) &= 0.0169 \cdot \Gamma_X^2 - 0.0573 \cdot \Gamma_X + 0.047\\
    &b_s(\theta) &= 2.529 \cdot 10^{-6} \cdot \theta^3 - 2.097 \cdot 10^{-4} \cdot \theta^2 + 2.850 \cdot 10^{-3} \cdot \theta + 0.0596\\
\end{eqnarray*}

and in the case of $\alpha=1$:

\begin{eqnarray*}
    &a_s(h_\text{X}) &= 1.858 \cdot h_\text{X}^{0.0065} \\
    &a_s(M_8) &= 0.082 \cdot M_8 + 0.992 \\
    &a_s(\dot{m}) &= 1.055 \cdot \dot{m}^{0.0178} \\
    &a_s(L_X) &= 1.046 \cdot L_X^{0.0098} \\
    &a_s(\Gamma_X) &= 0.0237 \cdot \Gamma^2 - 0.0825 \cdot \Gamma + 1.071\\
    &a_s(\theta) &= 0.8675 \cdot cos^2(\theta) - 0.4364 \cdot cos(\theta) + 0.8252\\
\\
    &b_s(h_\text{X}) &= -2.781 \cdot 10^{-7} \cdot h_\text{X}^2 + 2.058 \cdot 10^{-4} \cdot h_\text{X} - 0.119 \\
    &b_s(\dot{m}) &= 0.0455 \cdot \dot{m}^2 - 0.0131 \cdot \dot{m} + 0.0005 \\
    &b_s(L_X) &= -0.109 \cdot L_X^{-0.008} + 0.1131 \\
    &b_s(\Gamma_X) &= 4.65 \cdot 10^{-4} \cdot \Gamma_X^2 - 0.00166 \cdot \Gamma_X + 0.00146\\
    &b_s(\theta) &= 3.271 \cdot 10^{-7} \cdot \theta^3 + 6.750 \cdot 10^{-5} \cdot \theta^2 - 4.847 \cdot 10^{-3} \cdot \theta + 0.0649\\
\end{eqnarray*}

The model normalization, $N$ (in units of $(erg/s/cm^2/\AA)^2$) depends on the physical parameters differently at long and short wavelengths. At $\lambda \le 1950$\AA, and  $\alpha=0$,  we found that:

\begin{equation}
\begin{split}    
    N(\lambda, \mu) =&  a_{N1}(h_\text{X}) \cdot a_{N1}(M_8) \cdot a_{N1} (\dot{m}) \cdot a_{N1} (L_X) \cdot a_{N1} (\Gamma_X) \cdot a_{N1} (\theta) \cdot \lambda_{1950}^{-[b_{N1} (h_\text{X}) + b_{N1} (\dot{m}) + b_{N1} (L_X) + b_{N1}  (\Gamma_X)]}  +\\
     &+ c_{N1}(h_\text{X}) \cdot c_{N1}(M_8) \cdot c_{N1} (\dot{m}) \cdot c_{N1} (L_X) \cdot c_{N1} (\Gamma_X)\cdot c_{N1} (\theta) ,  \qquad  \lambda_{1950} \le 1 \\ 
\end{split}    
\end{equation}

where:

\begin{eqnarray*}
    &a_{N1}(h_\text{X}) &= 2.287 \cdot 10^{-20} \cdot h_\text{X}^{1.552} \\
    &a_{N1}(M_8) &= 575.4 \cdot M_8 ^{2.76} \\
    &a_{N1}(\dot{m}) &= 4.002  \cdot \frac{1}{1 +(\frac{\dot{m}}{0.013})^{0.816}}\\
    &a_{N1}(L_X) &= 1.313  \cdot \frac{1}{1 +(\frac{L_X}{0.05})^{0.721}}\\
    &a_{N1}(\Gamma) &= 50.500 \cdot \Gamma^4 - 368.958 \cdot \Gamma^3 + 1010.363 \cdot \Gamma^2 - 1227.991 \cdot \Gamma + 559.187 \\
    &a_{N1}(\theta) &= 14.593  \cdot \frac{1}{1 +(\frac{cos(\theta)}{2.983})^{-1.920}}\\
\\
    &b_{N1}(h_\text{X}) &= 5.735 \cdot 10^{-5} \cdot h_\text{X}^2 - 1.4065 \cdot 10^{-2} \cdot h_\text{X} + 6.1296 \\
    &b_{N1}(M_8) &= - 0.2322 \cdot M_8^2 + 0.247 \cdot M_8 - 0.0224 \\
    &b_{N1}(\dot{m}) &= -2.08 \cdot\dot{m}^2 +1.347 \cdot \dot{m} - 0.0621 \\
    &b_{N1}(L_X) &= 5.82 \cdot L_X^{-0.0057} - 5.975 \\
    &b_{N1}(\Gamma_X) &= -0.118 \cdot \Gamma^2 + 0.382 \cdot \Gamma - 0.292 \\
\\
    &c_{N1}(h_\text{X}) &= 1.416 \cdot 10^{-20} \cdot h_\text{X}^{1.568} \\
    &c_{N1}(M_8) &= 54.02 \cdot M_8 ^{2.738} \\
    &c_{N1}(\dot{m}) &= 4.333  \cdot \frac{1}{1 +(\frac{\dot{m}}{0.011})^{0.795}}\\
    &c_{N1}(L_X) &= 1.285  \cdot \frac{1}{1 +(\frac{L_X}{0.057})^{0.721}}\\
    &c_{N1}(\Gamma) &= 51.803 \cdot \Gamma^4 - 378.636 \cdot \Gamma^3 + 1037.145 \cdot \Gamma^2 - 1260.698 \cdot \Gamma + 574.054 \\
    &c_{N1}(\theta) &= 1.633 \cdot \left [cos(\theta) \right ]^{1.840} \\
\end{eqnarray*}

for $\alpha=0$, and

\begin{eqnarray*}
    &a_{N1}(h_\text{X}) &= 6.953 \cdot 10^{-20} \cdot h_\text{X}^{1.379} \\
    &a_{N1}(M_8) &= 524.81 \cdot M_8 ^{2.72} \\
    &a_{N1}(\dot{m}) &= 1.847  \cdot \frac{1}{1 +(\frac{\dot{m}}{0.062})^{0.774}}\\
    &a_{N1}(L_X) &= 2.201  \cdot \frac{1}{1 +(\frac{L_X}{0.0077})^{0.700}}\\
    &a_{N1}(\Gamma) &= 27.564 \cdot \Gamma^4 - 200.162 \cdot \Gamma^3 + 546.632 \cdot \Gamma^2 - 664.611 \cdot \Gamma + 303.973 \\
    &a_{N1}(\theta) &= 18.088  \cdot \frac{1}{1 +(\frac{cos(\theta)}{3.357})^{-1.922}}\\
\\
    &b_{N1}(h_\text{X}) &= 3.74 \cdot 10^{-5} \cdot h_\text{X}^2 - 1.093 \cdot 10^{-2} \cdot h_\text{X} + 5.924 \\
    &b_{N1}(M_8) &= - 0.0317 \cdot M_8^2 - 0.0416 \cdot M_8 + 0.0045 \\
    &b_{N1}(\dot{m}) &= -0.76 \cdot\dot{m}^2 + 0.644 \cdot \dot{m} - 0.0303 \\
    &b_{N1}(L_X) &= 5.833 \cdot L_X^{0.00082} - 5.811 \\
    &b_{N1}(\Gamma_X) &= 0.0162 \cdot \Gamma^2 - 0.0805 \cdot \Gamma - 0.096 \\
\\
    &c_{N1}(h_\text{X}) &= 4.49 \cdot 10^{-20} \cdot h_\text{X}^{1.39} \\
    &c_{N1}(M_8) &= 495.45 \cdot M_8 ^{2.695} \\
    &c_{N1}(\dot{m}) &= 1.93  \cdot \frac{1}{1 +(\frac{\dot{m}}{0.055})^{0.763}}\\
    &c_{N1}(L_X) &= 2.178  \cdot \frac{1}{1 +(\frac{L_X}{0.0079})^{0.696}}\\
    &c_{N1}(\Gamma) &= 27.772 \cdot \Gamma^4 - 201.715 \cdot \Gamma^3 + 550.938 \cdot \Gamma^2 - 669.869 \cdot \Gamma + 306.357 \\
    &c_{N1}(\theta) &= 1.639 \cdot \left [cos(\theta) \right ]^{1.855} \\
\end{eqnarray*}

for $\alpha=1$. For $\lambda > 1950$\AA, and $\alpha=0$, we find: 

\begin{equation}
    N(\lambda, \mu) =  a_{N2}(h_\text{X}) \cdot a_{N2}(M_8) \cdot a_{N2} (\dot{m}) \cdot a_{N2} (L_X) \cdot a_{N2} (\Gamma_X) \cdot a_{N2} (\theta) \cdot \lambda_{1950}^{-[b_{N2} (h_\text{X}) + b_{N2} (\dot{m}) + b_{N2} (L_X) + b_{N2}  (\Gamma_X)]},  \qquad  \lambda_{1950} > 1  \\
\end{equation}

where:

\begin{eqnarray*}
    &a_{N2}(h_\text{X}) &= 3.6087 \cdot 10^{-20} \cdot h_\text{X}^{1.57} \\
    &a_{N2}(M_8) &= 549.54 \cdot M_8 ^{2.74} \\
    &a_{N2}(\dot{m}) &= 4.292  \cdot \frac{1}{1 +(\frac{\dot{m}}{0.011})^{0.787}}\\
    &a_{N2}(L_X) &= 1.303  \cdot \frac{1}{1 +(\frac{L_X}{0.054})^{0.708}}\\
    &a_{N2}(\Gamma) &= 51.474 \cdot \Gamma^4 - 376.181 \cdot \Gamma^3 + 1030.331 \cdot \Gamma^2 - 1252.363 \cdot \Gamma + 570.265 \\
    &a_{N2}(\theta) &= 1.641 \cdot \left [cos(\theta) \right ]^{1.857} \\
\\
    &b_{N2}(h_\text{X}) &= -2.938 \cdot 10^{-3} \cdot h_\text{X} + 4.897 \\
    &b_{N2}(M_8) &= 0.2537 \cdot M_8^2 - 0.2087 \cdot M_8 + 0.0183 \\
    &b_{N2}(\dot{m}) &= 0.346 \cdot \dot{m} - 0.017 \\
    &b_{N2}(L_X) &= -9.893 \cdot L_X^3 + 8.229 \cdot L_X^2 - 1.635 \cdot L_X + 0.016 \\
    &b_{N2}(\Gamma_X) &= -0.0739 \cdot \Gamma^2 + 0.249 \cdot \Gamma - 0.202 \\
\end{eqnarray*}

For $\lambda > 1950$\AA, and $\alpha=1$:

\begin{eqnarray*}
    &a_{N2}(h_\text{X}) &= 1.13 \cdot 10^{-19} \cdot h_\text{X}^{1.392} \\
    &a_{N2}(M_8) &= 512.86 \cdot M_8 ^{2.71} \\
    &a_{N2}(\dot{m}) &= 1.892  \cdot \frac{1}{1 +(\frac{\dot{m}}{0.058})^{0.767}}\\
    &a_{N2}(L_X) &= 2.231  \cdot \frac{1}{1 +(\frac{L_X}{0.0074})^{0.689}}\\
    &a_{N2}(\Gamma) &= 27.675 \cdot \Gamma^4 - 200.995 \cdot \Gamma^3 + 548.946 \cdot \Gamma^2 - 667.439 \cdot \Gamma + 305.254 \\
    &a_{N2}(\theta) &= 1.646 \cdot \left [cos(\theta) \right ]^{1.870} \\
\\
    &b_{N2}(h_\text{X}) &= -2.529 \cdot 10^{-3} \cdot h_\text{X} + 4.818 \\
    &b_{N2}(M_8) &= -0.0056 \cdot M_8^2 + 0.0429 \cdot M_8 - 0.0042 \\
    &b_{N2}(\dot{m}) &= 0.094 \cdot \dot{m} - 0.005 \\
    &b_{N2}(L_X) &= -5.863 \cdot L_X^3 + 4.823 \cdot L_X^2 - 0.832 \cdot L_X + 0.008 \\
    &b_{N2}(\Gamma_X) &= -0.029 \cdot \Gamma^2 + 0.093 \cdot \Gamma - 0.07 \\
\end{eqnarray*}

The above equations provide a good approximation to the transfer functions, as shown by the residual plots in the figures we plot in the following Section. Nonetheless, the agreement between $|\Gamma_{approx}(\nu)|^2$ and $|\Gamma_\lambda(\nu)|^2$  is not perfect. We show below that this is not a problem when we fit the observed PSDs. 

Firstly, we study the case when we approximate $|\Gamma_\lambda(\nu)|^2$ using eq. \ref{eq:transfer_benpl} in a single waveband. As an example, we considered the   $|\Gamma_\lambda(\nu)|^2$ functions plotted in Fig.\, \ref{fig:transfer_ratio_height}, for the eight  X-ray corona heights we considered. For each one of them we recorded $|\Gamma_\lambda(\nu)|^2$ at ten distinct frequencies, which are comparable to the frequencies at which the NGC 5548 PSDs  were calculated. We have, thus, created eight sets of ten ($\nu$, $|\Gamma(\nu)|^2$) points, and we fit them\footnote{The fit was conducted using the \textit{lmfit} package (https://lmfit.github.io/lmfit-py/) in the \textit{python} programming language.} with eq.\, \ref{eq:transfer_benpl}.

\begin{figure}
\includegraphics[width=0.47\linewidth,height=0.47\linewidth, trim={50 0 50 20}, clip]{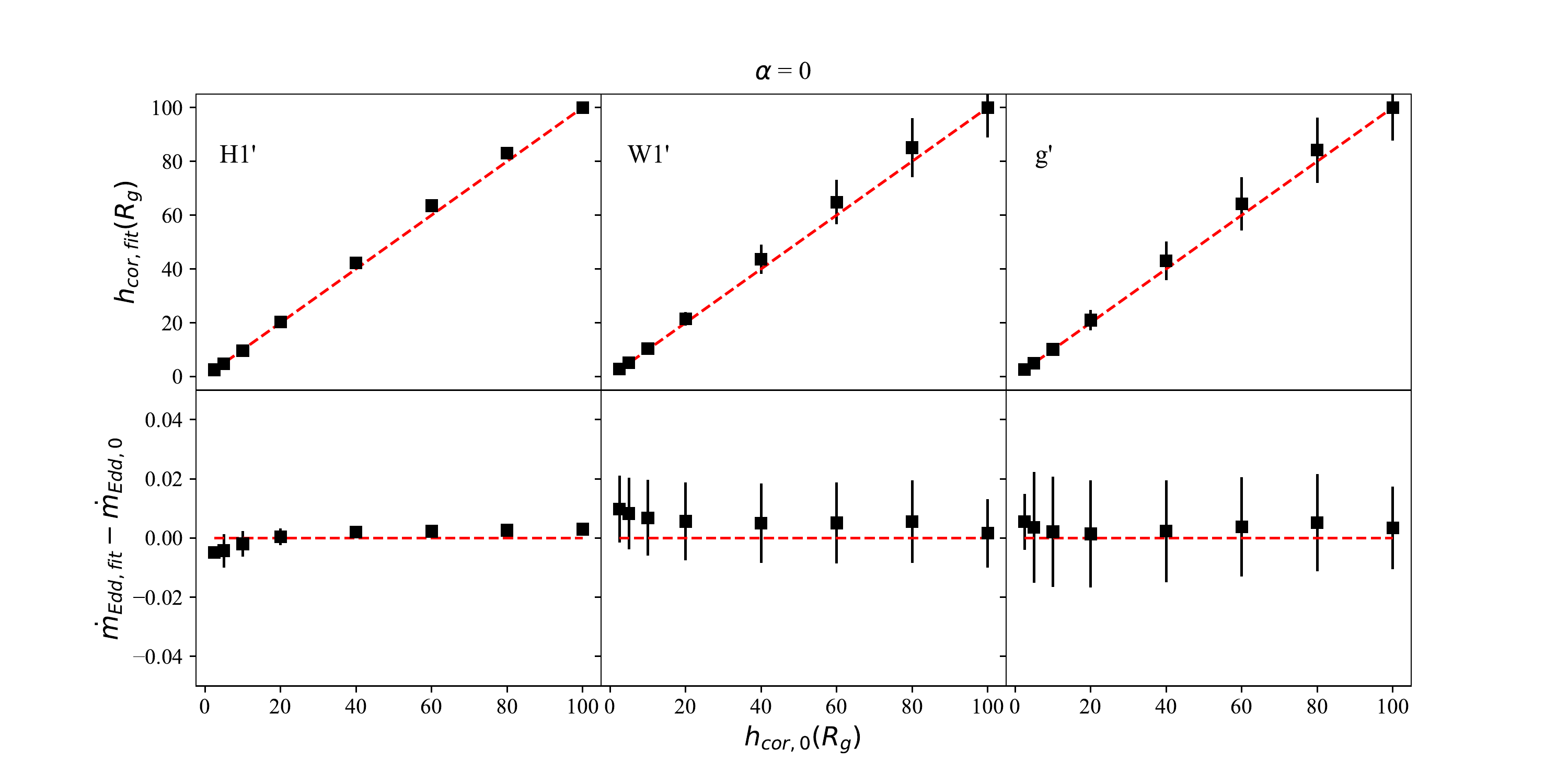}
\includegraphics[width=0.47\linewidth,height=0.47\linewidth, trim={50 0 50 20}, clip]{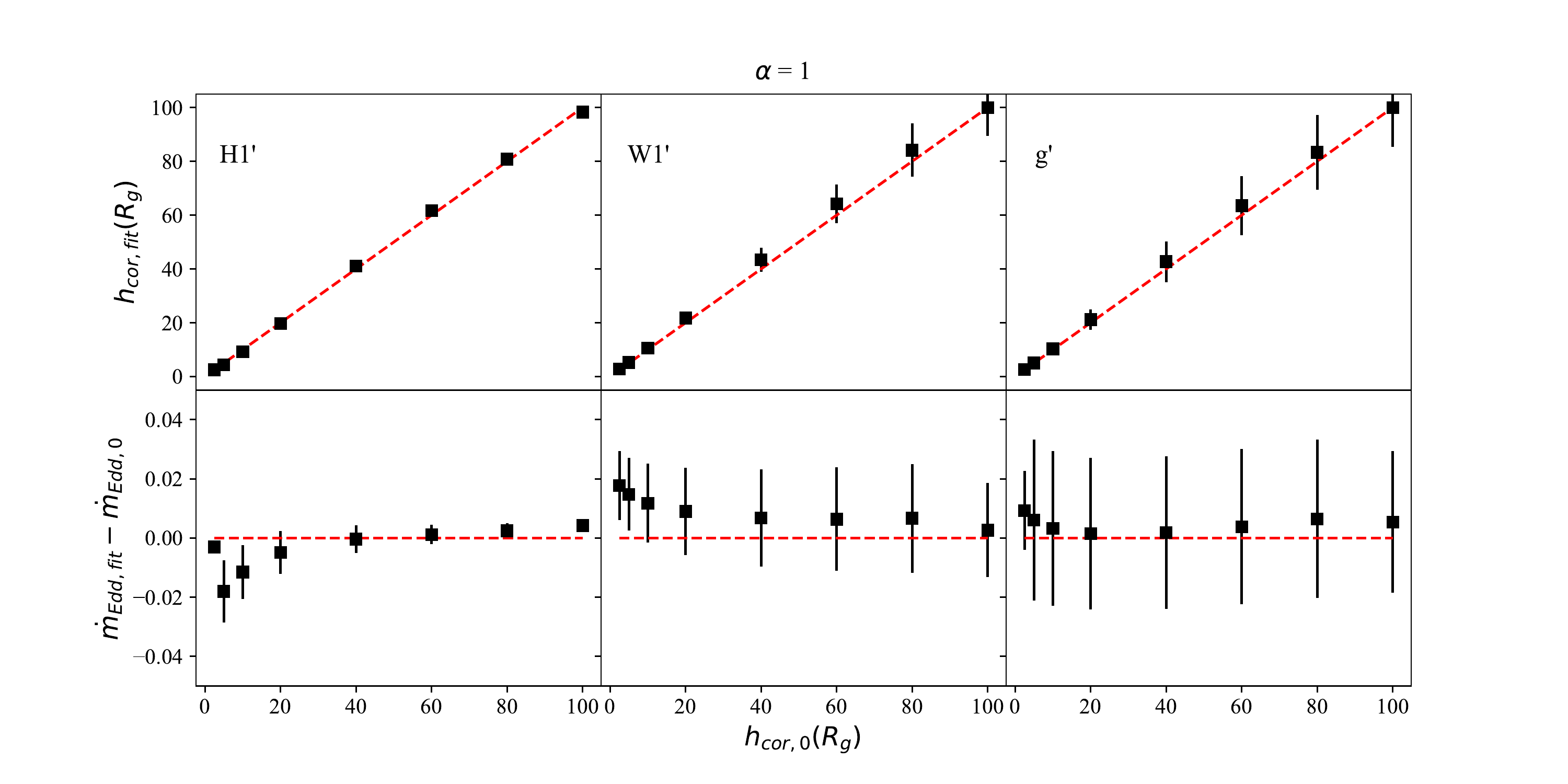}
\caption{\textit{Upper}: The best-fit corona height, obtained when the test $|\Gamma_\lambda(\nu)|^2$ are fit, versus the input heights used in estimating the simulated $|\Gamma_\lambda(\nu)|^2$. The red dashed line indicates the identity line. The plotted results correspond to the case when each of the wavebands H1', W1', and g' are considered individually. \textit{Lower}: The difference between the best fit value and the input value of the accretion rate, for the different heights. The plotted errors correspond to the 3-$\sigma$ level. The left plots present the results for a black hole with zero spin, $\alpha = 0$, and the right plots correspond to $\alpha=1$.} 
\label{fig:asses_1band}
\end{figure}

The results are plotted in Fig.\ref{fig:asses_1band}, in the case when we let the corona height and the accretion rate to vary during the fit, while all the other parameters are fixed to the their intrinsic values. The upper panel plots in this figure show the corona height determined by the fit versus the intrinsic $h_\text{X}$, whereas the lower panel plots show the difference between the best-fit and the intrinsic accretion rate, for different values of the input $h_\text{X}$. Evidently, the model fits can retrieve the input values with high accuracy, in all wavebands. We have confirmed that the result is unchanged when we let a different parameter, instead of the accretion rate, to be minimized during the fit. However, $h_\text{X}$ and $\dot{m}$ will probably be the parameters of interest in most applications, as the other model parameters are usually independently known (or assumed).  When we increase the number of model parameters that we let free, then their error increases, as expected. We reach the same results when we consider the transfer functions for different BH masses, accretion rates etc.   

\begin{figure}
\includegraphics[width=0.48\linewidth,height=0.5\linewidth, trim={0 0 0 0}, clip]{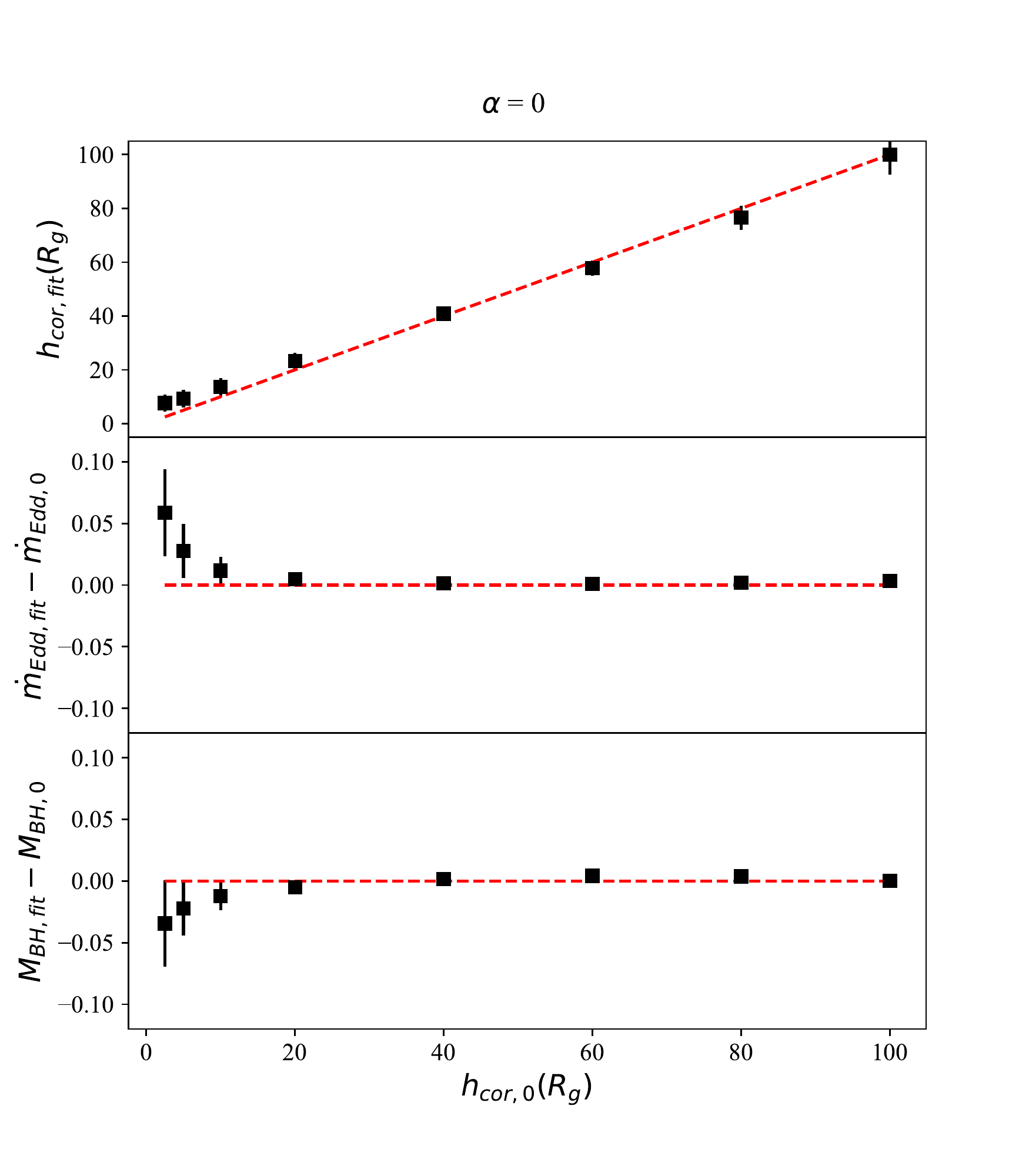}
\hspace{0.05\linewidth}
\includegraphics[width=0.48\linewidth,height=0.5\linewidth, trim={0 0 0 0}, clip]{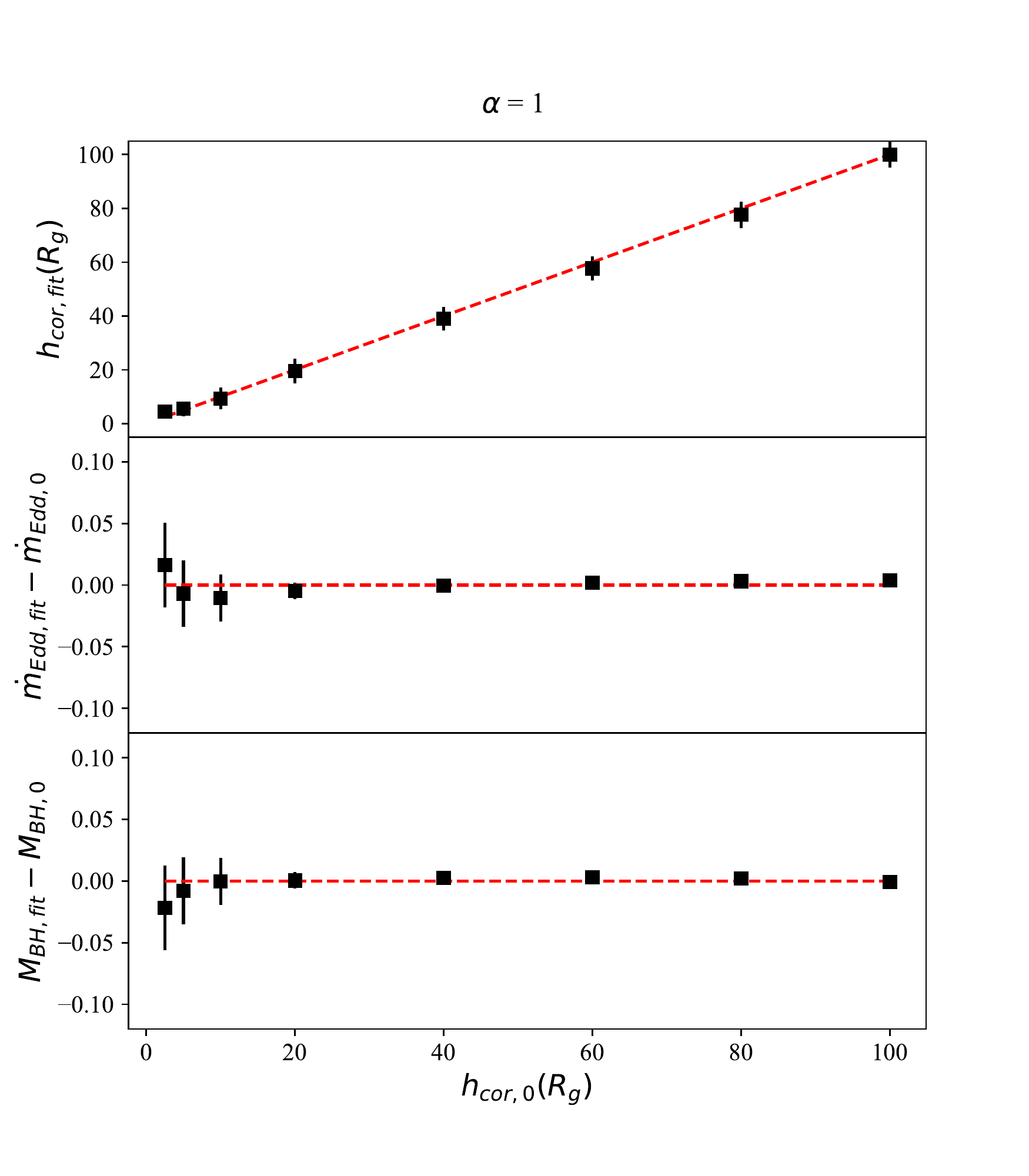}
\caption{Best-fit results obtained when eq. \ref{eq:transfer_benpl} is used to reproduce simulated $|\Gamma_\lambda(\nu)|^2$ produced for different heights of the X-ray source. The best-fit corona height (upper) as well as the difference between the best-fit and the input values for the accretion rate (middle) and the black hole mass (bottom panel) for the various input values of $h_\text{X}$ are plotted. The left plots present the results for a zero spin black hole, while the right plots correspond to $\alpha=1$. The plotted errors denote to the 3-$\sigma$ confidence interval.} 
\label{fig:asses_5bands}
\end{figure}

Then we repeated the above experiment to test whether eq. \ref{eq:transfer_benpl} can be used to determine, accurately, any combination of three parameters, when we fit $|\Gamma(\nu)|^2$ with eq.\,\ref{eq:transfer_benpl} at five wavebands, simultaneously. In this case, we considered the wavebands H1', W2', W1', g', and r'. The results are plotted in Fig. \ref{fig:asses_5bands}, in the case when the corona height, the accretion rate, and the black hole mass were assumed unknown and left free during the fit. The best-fit values are indeed in agreement with the input values.

The analysis above demonstrates that, in the case of the model fits to observed PSDs, eq.\, \ref{eq:transfer_benpl}, together with all the other equations we list above, can approximate very well the intrinsic transfer functions. Since we fit the NGC 5548 PSDs in five wavebands, and we let only two parameters free (height and accretion rate), we are confident that the best fit parameters we get are not affected, in any way, by the fact that we must use $|\Gamma_{approx, \lambda}(\nu)|^2$ instead of $|\Gamma_\lambda(\nu)|^2$ in the fits. In fact, our result show that, as long as the PSDs is measured in at least 5 wavebands, model fitting based on the use of $|\Gamma_{approx, \lambda}(\nu)|^2$ can determine, accurately, up to three model parameters.

\newpage

\section{Plots of the transfer function}
\label{sec:plots}

Plots of $|\Gamma(\nu)|^2$ vs frequency, up to the limiting frequency $\nu_\text{lim}=5 \cdot 10^{-5} \text{Hz} = 4.32 \text{day}^{-1}$ (see \S\ref{sec:approximation}), for all the physical parameters we consider. The lower panels  show the ratio of $|\Gamma_\lambda(\nu)|^2$ over $|\Gamma_{approx, \lambda}(\nu)^2|$. The latter function provides a good representation of $|\Gamma_\lambda(\nu)|^2$, within $\sim 5$\% in the UV, to $\sim 20$\%, at longer wavelengths. Larger amplitude residuals appear at high frequencies, for the largest (and smallest) parameter values. 

\begin{figure}[h]
\includegraphics[width=0.49\linewidth,height=0.45\linewidth, trim={50 10 90 50}, clip]{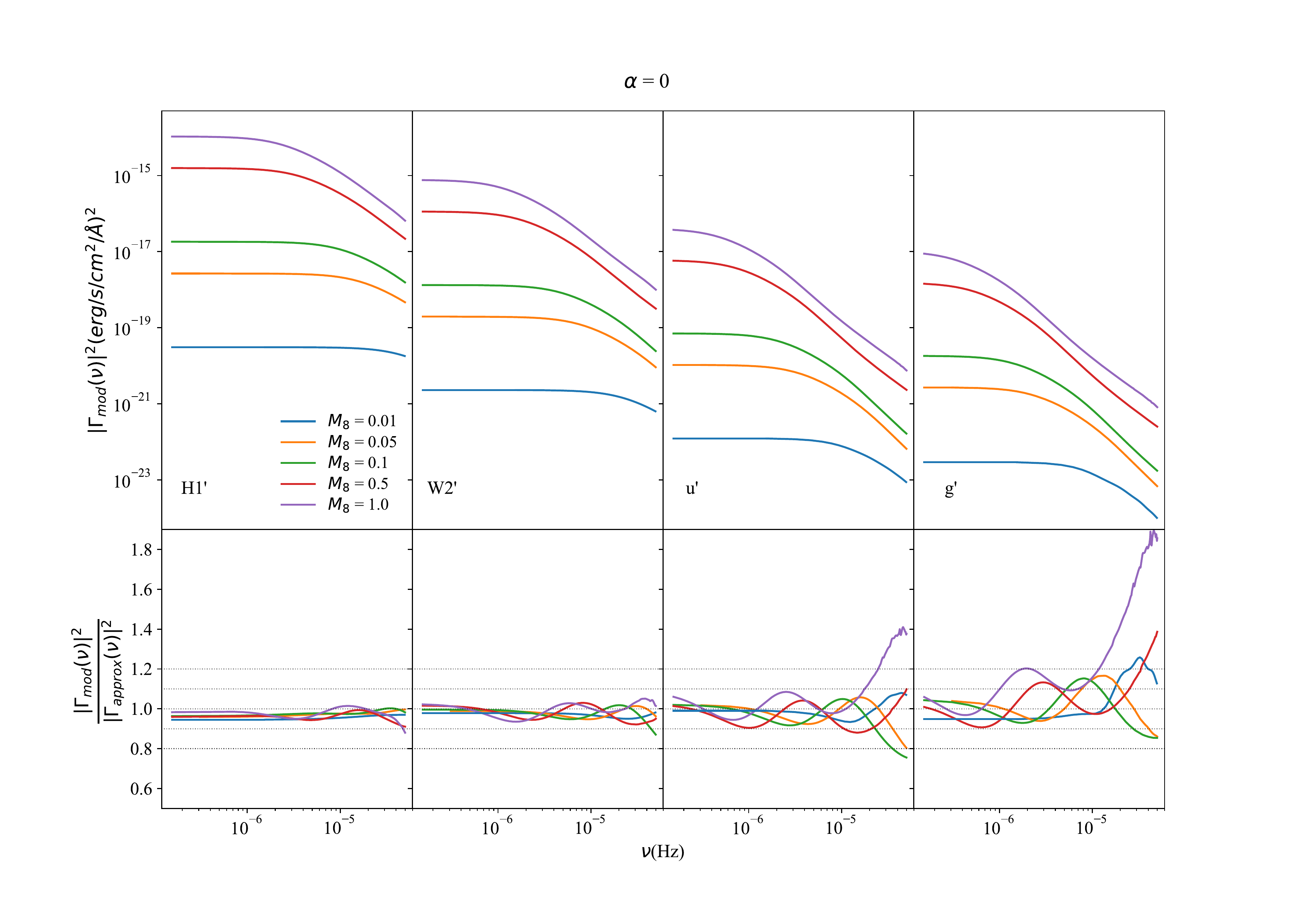}
\includegraphics[width=0.49\linewidth,height=0.45\linewidth, trim={50 10 90 50}, clip]{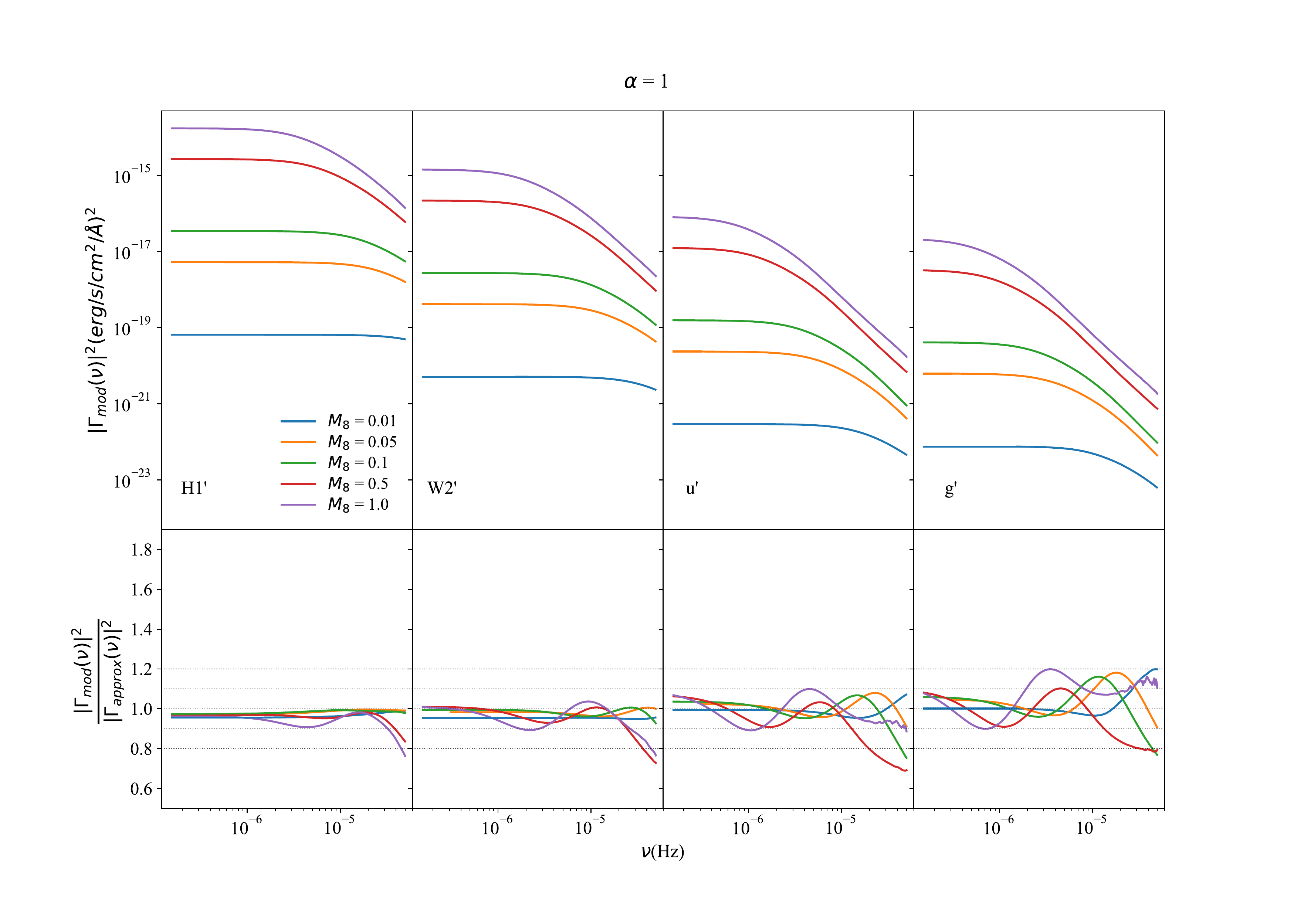}
\caption{\textit{Upper panel}: The modulus squared of the transfer function for spin 0 (left) and 1 (right), for different black hole masses, at  four wavebands (see Table \ref{tab:wavebands_details}). $|\Gamma_\lambda(\nu)|^2$ are computed using the K21 response functions and eq. \ref{eq:transfer_compute}. They are plotted up to $\nu_{\rm lim}$ {\bf (see \S\ref{sec:approximation}}). \textit{Bottom}: The ratio between $|\Gamma_\lambda(\nu)|^2$ over $|\Gamma_{approx}(\nu)|^2$, as defined by eq. \ref{eq:transfer_benpl}, using the parametric equations presented in Appendix \ref{sec:a1}. The dotted gray lines show the horizontal lines $y=1.2, 1.1, 1.0, 0.9, 0.8$, to help quantify how well $|\Gamma_{approx}(\nu)|^2$ approximates $|\Gamma_\lambda(\nu)|^2$. }    
\label{fig:transfer_ratio_mass}
\end{figure}

\begin{figure}[h]
\includegraphics[width=0.49\linewidth,height=0.45\linewidth, trim={50 10 90 50}, clip]{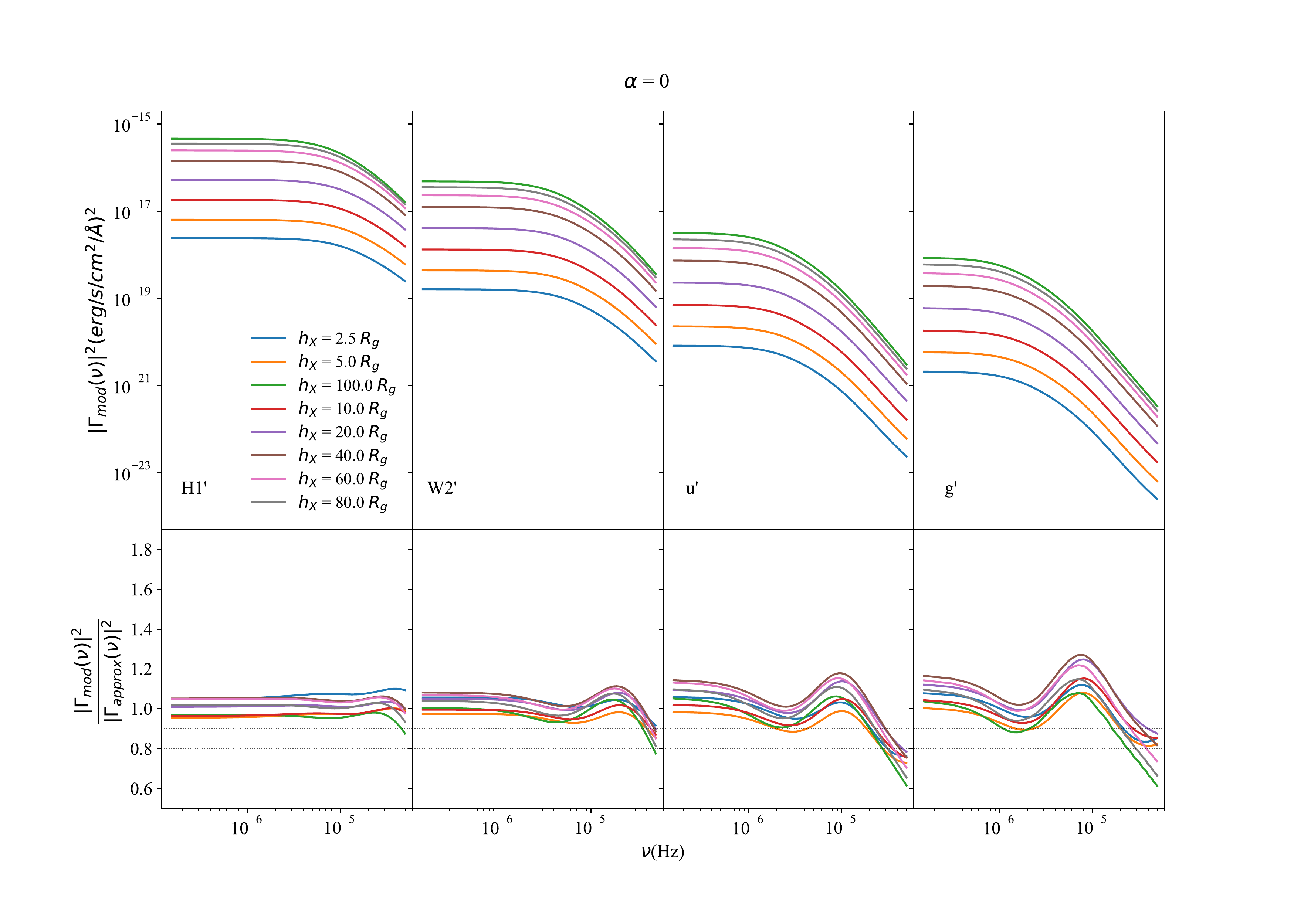}
\includegraphics[width=0.49\linewidth,height=0.45\linewidth, trim={50 10 90 50}, clip]{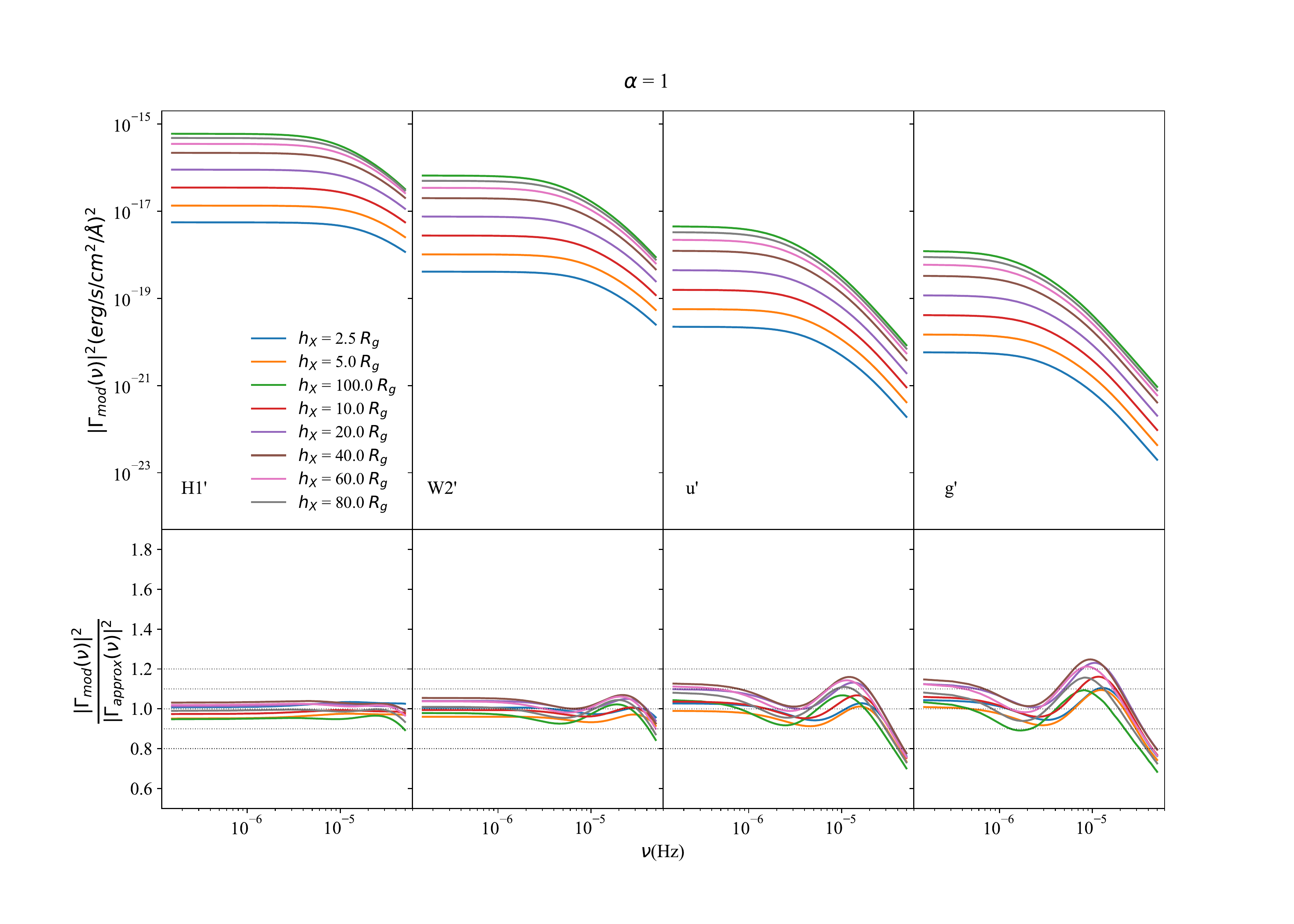}
\caption{Same as in Fig. \ref{fig:transfer_ratio_mass} for different values of the X-ray source height.}   \label{fig:transfer_ratio_height}
\end{figure}

\begin{figure}[h]
\includegraphics[width=0.49\linewidth,height=0.45\linewidth, trim={50 10 90 50}, clip]{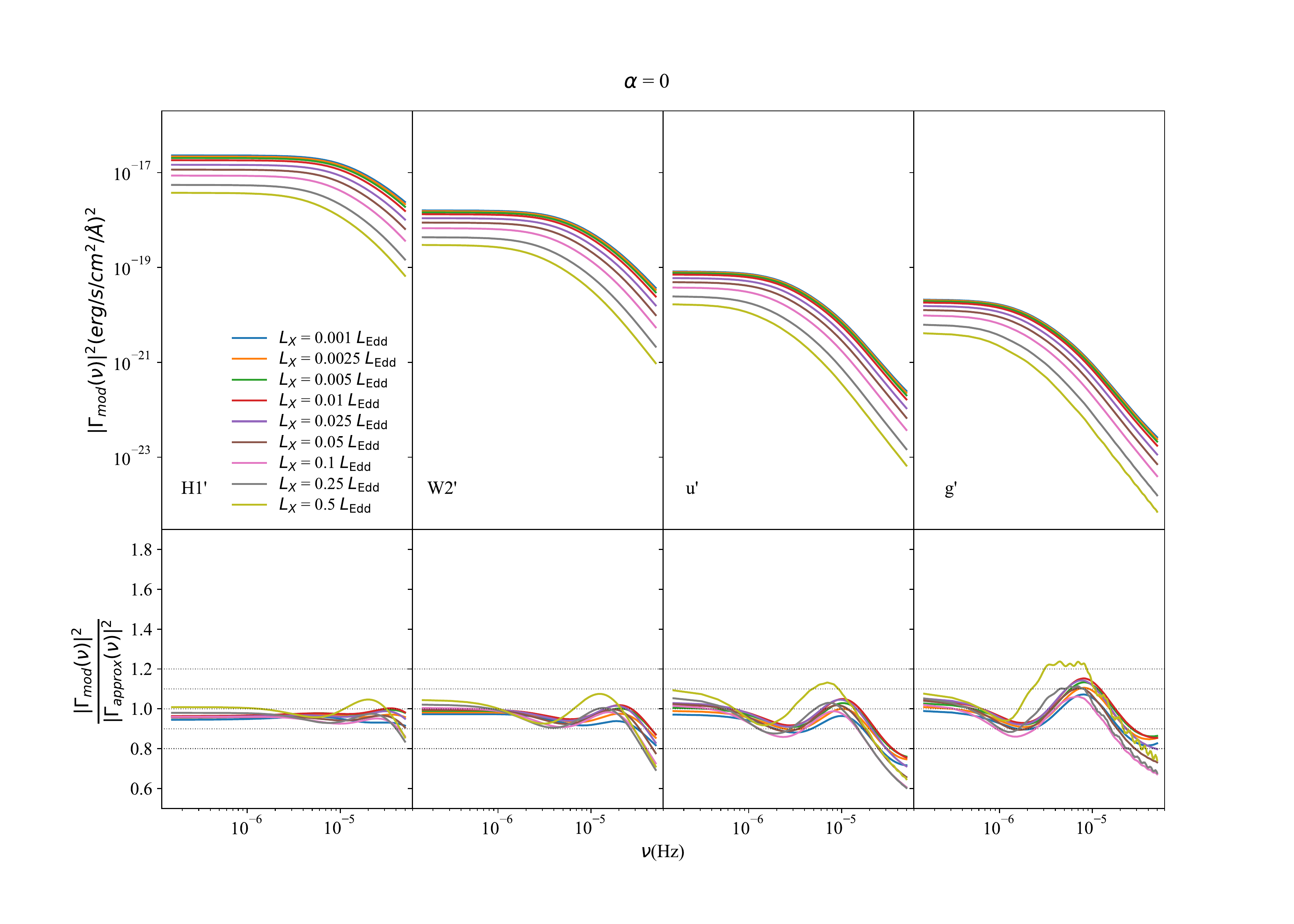}
\includegraphics[width=0.49\linewidth,height=0.45\linewidth, trim={50 10 90 50}, clip]{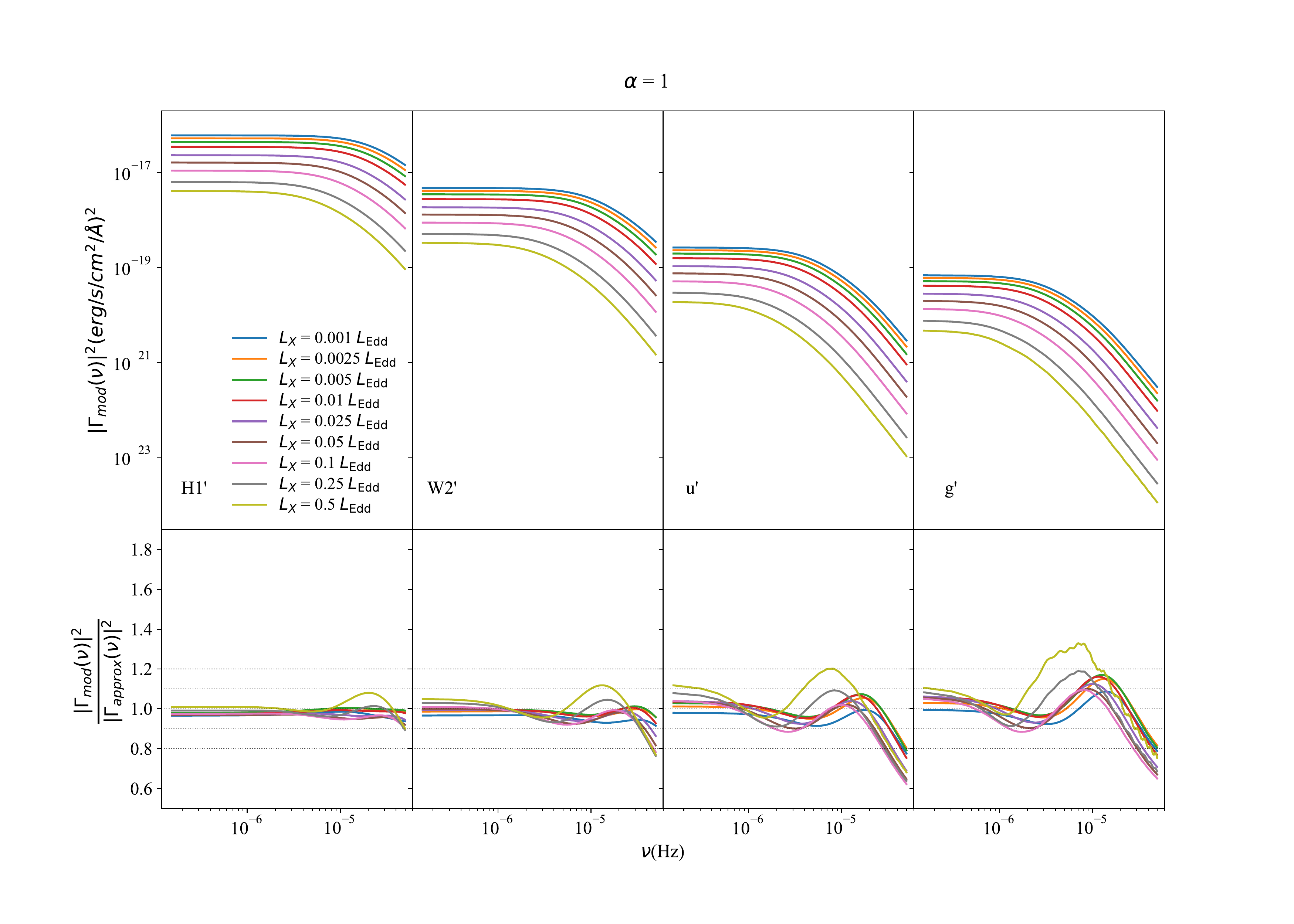}
\caption{Same as in Fig. \ref{fig:transfer_ratio_mass} for different values of the X-ray luminosity.}   \label{fig:transfer_ratio_lumin}
\end{figure}

\begin{figure}[h]
\includegraphics[width=0.49\linewidth,height=0.45\linewidth, trim={50 10 90 50}, clip]{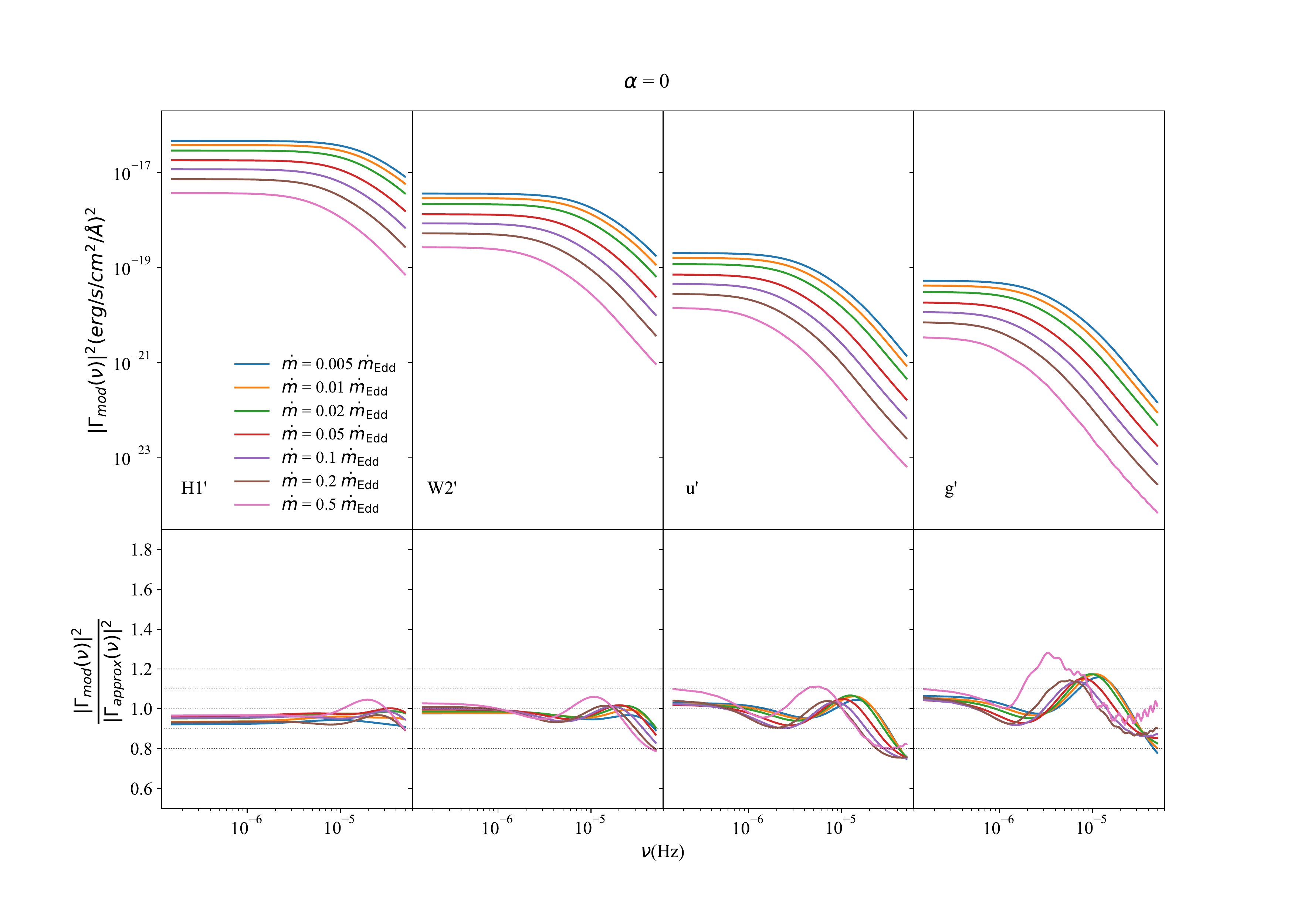}
\includegraphics[width=0.49\linewidth,height=0.45\linewidth, trim={50 10 90 50}, clip]{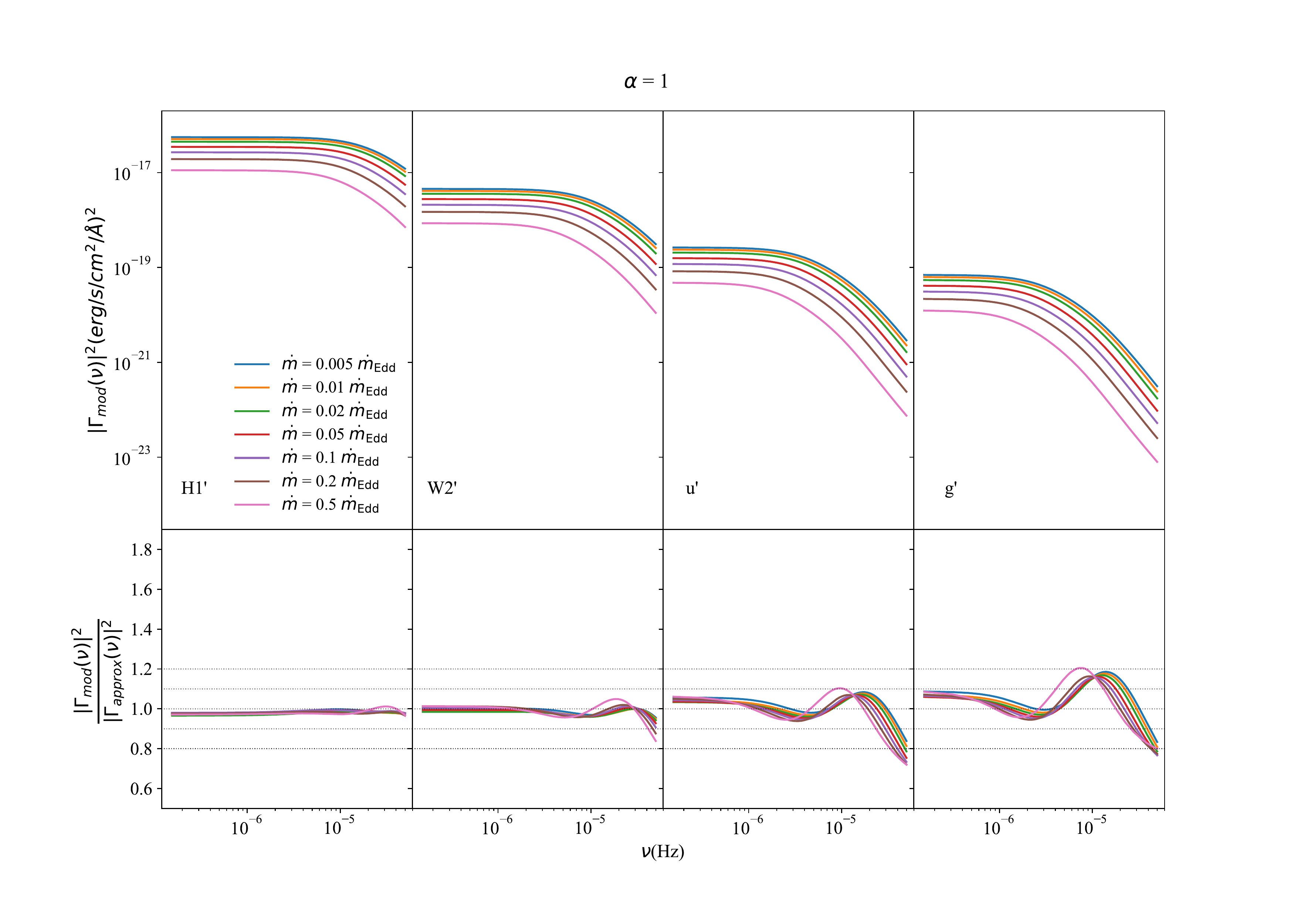}
\caption{Same as in Fig. \ref{fig:transfer_ratio_mass} for different values of the accretion rate.}   
\label{fig:transfer_ratio_mdot}
\end{figure}

\begin{figure}[h]
\includegraphics[width=0.49\linewidth,height=0.45\linewidth, trim={50 10 90 50}, clip]{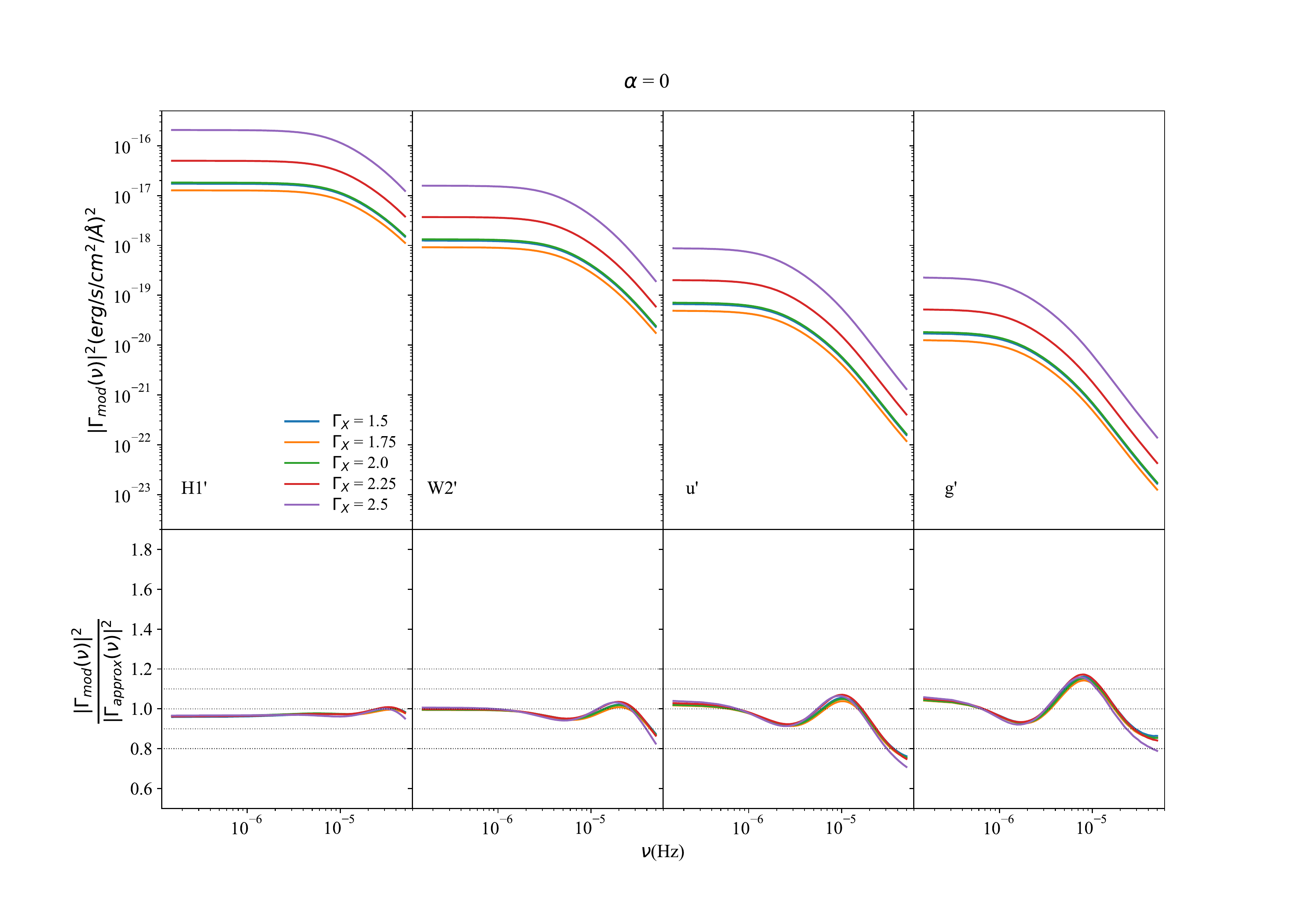}
\includegraphics[width=0.49\linewidth,height=0.45\linewidth, trim={50 10 90 50}, clip]{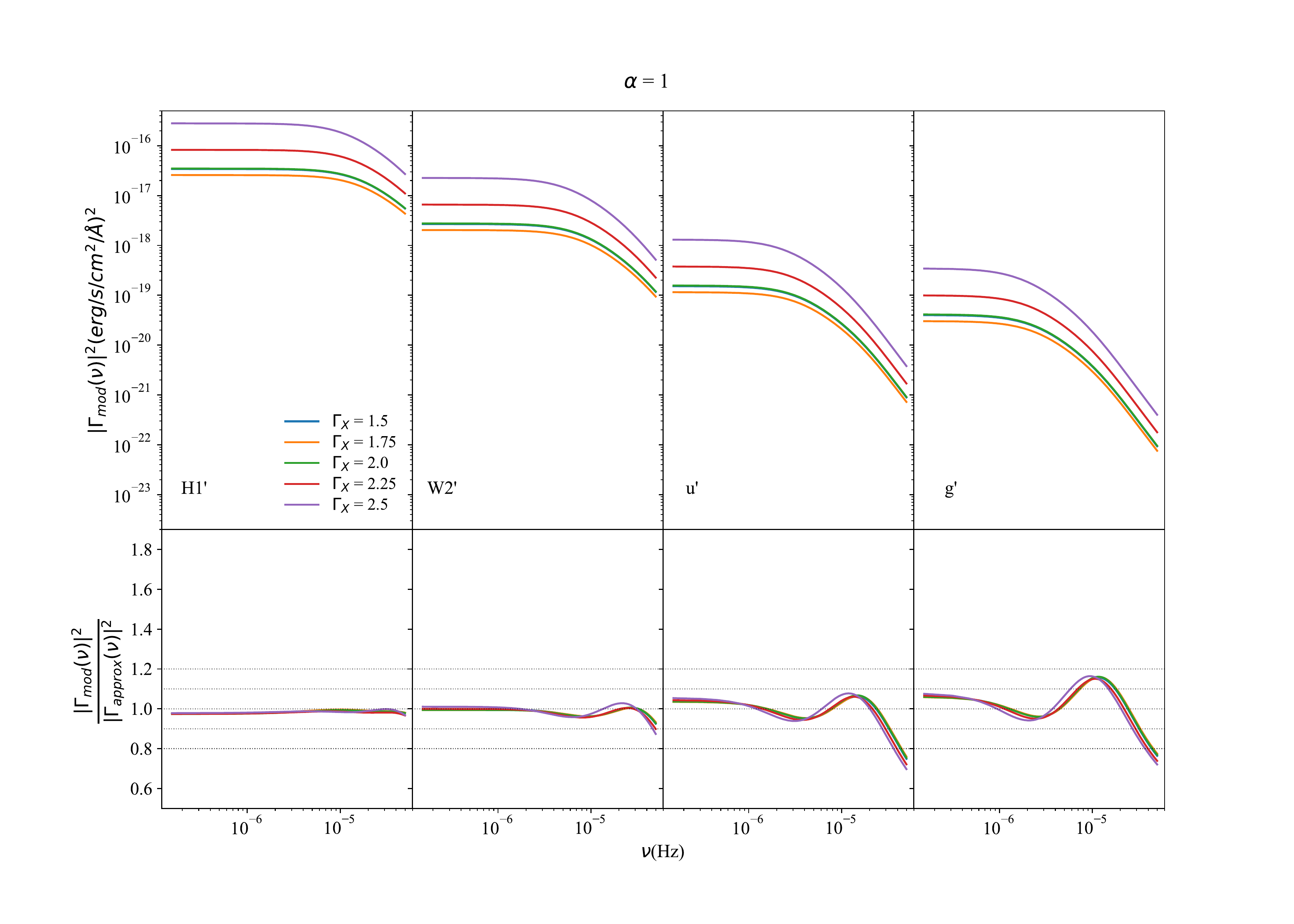}
\caption{Same as in Fig. \ref{fig:transfer_ratio_mass} for different values of the X-ray photon index. }   \label{fig:transfer_ratio_gamma}
\end{figure}

\begin{figure}[h]
\includegraphics[width=0.49\linewidth,height=0.45\linewidth, trim={50 10 90 50}, clip]{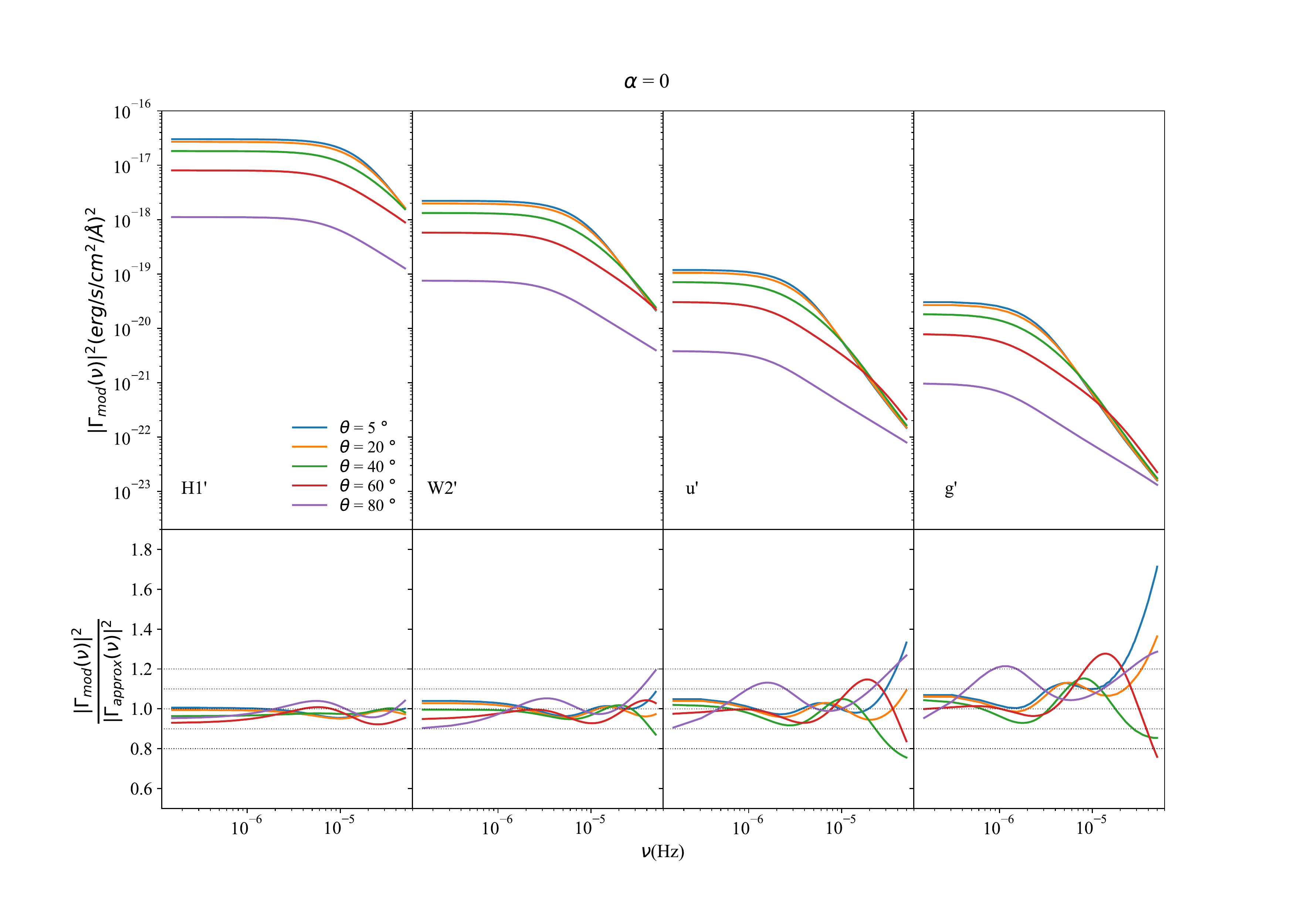}
\includegraphics[width=0.49\linewidth,height=0.45\linewidth, trim={50 10 90 50}, clip]{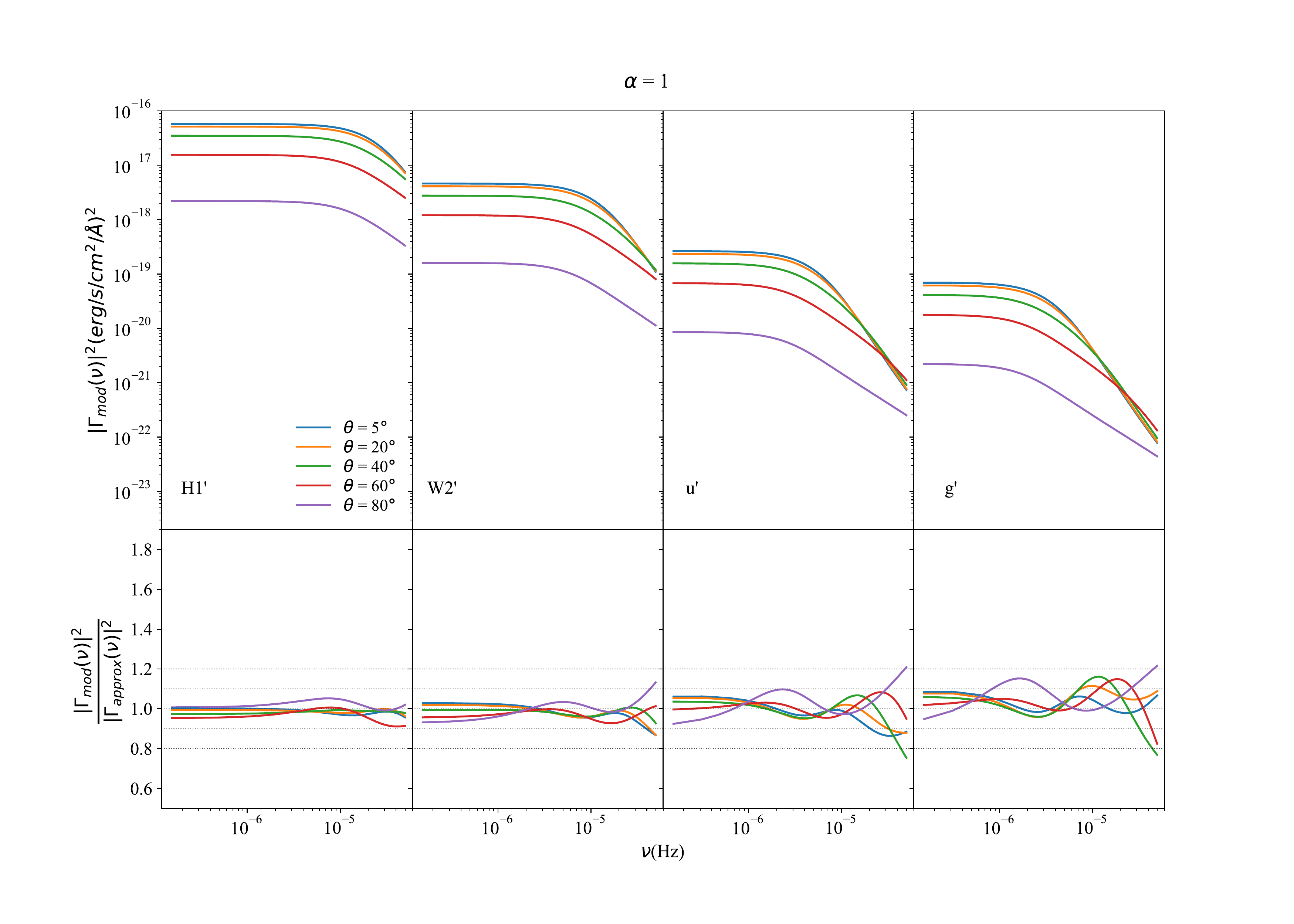}
\caption{Same as in Fig. \ref{fig:transfer_ratio_mass} for different values of system inclination.}     
\label{fig:transfer_ratio_incl}
\end{figure}



\end{document}